\documentclass{JHEP}
\usepackage{amssymb}

\preprint{{\tt  DAMTP-2005-46, hep-th/0504216}}

\title{Construction of one-loop ${\cal N}=4$ SYM effective action in the
harmonic superspace approach}

\author{I.L. Buchbinder\\ Department of Applied Mathematics and Theoretical Physics\\
University of Cambridge, Centre for Mathematical Sciences\\
Wilberforce Road, Cambridge, CB3 0WA, UK\\ and\\ Department of
Theoretical Physics\\ Tomsk State Pedagogical University\\ Tomsk
634041, Russia \\\email{joseph@tspu.edu.ru}}

\author{N.G. Pletnev\\ Department of Theoretical Physics\\ Institute of
Mathematics
\\ Novosibirsk 630090, Russia\\
\email{pletnev@math.nsc.ru}}

\immediate\write16{<<WARNING: LINEDRAW macros work with
emTeX-dvivers
                    and other drivers supporting emTeX \special's
                    (dviscr, dvihplj, dvidot, dvips, dviwin, etc.) >>}
\newdimen\Lengthunit       \Lengthunit  = 1.5cm
\newcount\Nhalfperiods     \Nhalfperiods= 9
\newcount\magnitude        \magnitude = 1000

\catcode`\*=11
\newdimen\L*   \newdimen\d*   \newdimen\d**
\newdimen\dm*  \newdimen\dd*  \newdimen\dt*
\newdimen\a*   \newdimen\b*   \newdimen\c*
\newdimen\a**  \newdimen\b**
\newdimen\xL*  \newdimen\yL*
\newdimen\rx*  \newdimen\ry*
\newdimen\tmp* \newdimen\linwid*

\newcount\k*   \newcount\l*   \newcount\m*
\newcount\k**  \newcount\l**  \newcount\m**
\newcount\n*   \newcount\dn*  \newcount\r*
\newcount\N*   \newcount\*one \newcount\*two  \*one=1 \*two=2
\newcount\*ths \*ths=1000
\newcount\angle*  \newcount\q*  \newcount\q**
\newcount\angle** \angle**=0
\newcount\sc*     \sc*=0

\newtoks\cos*  \cos*={1}
\newtoks\sin*  \sin*={0}

\catcode`\[=13

\def\rotate(#1){\advance\angle**#1\angle*=\angle**
\q**=\angle*\ifnum\q**<0\q**=-\q**\fi
\ifnum\q**>360\q*=\angle*\divide\q*360\multiply\q*360\advance\angle*-\q*\fi
\ifnum\angle*<0\advance\angle*360\fi\q**=\angle*\divide\q**90\q**=\q**
\def\sgcos*{+}\def\sgsin*{+}\relax
\ifcase\q**\or
 \def\sgcos*{-}\def\sgsin*{+}\or
 \def\sgcos*{-}\def\sgsin*{-}\or
 \def\sgcos*{+}\def\sgsin*{-}\else\fi
\q*=\q** \multiply\q*90\advance\angle*-\q*
\ifnum\angle*>45\sc*=1\angle*=-\angle*\advance\angle*90\else\sc*=0\fi
\def[##1,##2]{\ifnum\sc*=0\relax
\edef\cs*{\sgcos*.##1}\edef\sn*{\sgsin*.##2}\ifcase\q**\or
 \edef\cs*{\sgcos*.##2}\edef\sn*{\sgsin*.##1}\or
 \edef\cs*{\sgcos*.##1}\edef\sn*{\sgsin*.##2}\or
 \edef\cs*{\sgcos*.##2}\edef\sn*{\sgsin*.##1}\else\fi\else
\edef\cs*{\sgcos*.##2}\edef\sn*{\sgsin*.##1}\ifcase\q**\or
 \edef\cs*{\sgcos*.##1}\edef\sn*{\sgsin*.##2}\or
 \edef\cs*{\sgcos*.##2}\edef\sn*{\sgsin*.##1}\or
 \edef\cs*{\sgcos*.##1}\edef\sn*{\sgsin*.##2}\else\fi\fi
\cos*={\cs*}\sin*={\sn*}\global\edef\gcos*{\cs*}\global\edef\gsin*{\sn*}}\relax
\ifcase\angle*[9999,0]\or
[999,017]\or[999,034]\or[998,052]\or[997,069]\or[996,087]\or
[994,104]\or[992,121]\or[990,139]\or[987,156]\or[984,173]\or
[981,190]\or[978,207]\or[974,224]\or[970,241]\or[965,258]\or
[961,275]\or[956,292]\or[951,309]\or[945,325]\or[939,342]\or
[933,358]\or[927,374]\or[920,390]\or[913,406]\or[906,422]\or
[898,438]\or[891,453]\or[882,469]\or[874,484]\or[866,499]\or
[857,515]\or[848,529]\or[838,544]\or[829,559]\or[819,573]\or
[809,587]\or[798,601]\or[788,615]\or[777,629]\or[766,642]\or
[754,656]\or[743,669]\or[731,681]\or[719,694]\or[707,707]\or
\else[9999,0]\fi}

\catcode`\[=12

\def\GRAPH(hsize=#1)#2{\hbox to #1\Lengthunit{#2\hss}}

\def\Linewidth#1{\global\linwid*=#1\relax
\global\divide\linwid*10\global\multiply\linwid*\mag
\global\divide\linwid*100\special{em:linewidth \the\linwid*}}

\Linewidth{.4pt}
\def\sm*{\special{em:moveto}}
\def\sl*{\special{em:lineto}}
*
*
\newbox\spm*   \newbox\spl*
\setbox\spm*\hbox{\sm*} \setbox\spl*\hbox{\sl*}

\def\mov#1(#2,#3)#4{\rlap{\L*=#1\Lengthunit
\xL*=#2\L* \yL*=#3\L* \xL*=\xscale\xL* \yL*=\yscale\yL* \rx*
\the\cos*\xL* \tmp* \the\sin*\yL* \advance\rx*-\tmp* \ry*
\the\cos*\yL* \tmp* \the\sin*\xL* \advance\ry*\tmp*
\kern\rx*\raise\ry*\hbox{#4}}}

\def\rmov*(#1,#2)#3{\rlap{\xL*=#1\yL*=#2\relax
\rx* \the\cos*\xL* \tmp* \the\sin*\yL* \advance\rx*-\tmp* \ry*
\the\cos*\yL* \tmp* \the\sin*\xL* \advance\ry*\tmp*
\kern\rx*\raise\ry*\hbox{#3}}}

\def\lin#1(#2,#3){\rlap{\sm*\mov#1(#2,#3){\sl*}}}

\def\arr*(#1,#2,#3){\rmov*(#1\dd*,#1\dt*){\sm*
\rmov*(#2\dd*,#2\dt*){\rmov*(#3\dt*,-#3\dd*){\sl*}}\sm*
\rmov*(#2\dd*,#2\dt*){\rmov*(-#3\dt*,#3\dd*){\sl*}}}}

\def\arrow#1(#2,#3){\rlap{\lin#1(#2,#3)\mov#1(#2,#3){\relax
\d**=-.012\Lengthunit\dd*=#2\d**\dt*=#3\d**
\arr*(1,10,4)\arr*(3,8,4)\arr*(4.8,4.2,3)}}}

\def\arrlin#1(#2,#3){\rlap{\L*=#1\Lengthunit\L*=.5\L*
\lin#1(#2,#3)\rmov*(#2\L*,#3\L*){\arrow.1(#2,#3)}}}

\def\dasharrow#1(#2,#3){\rlap{{\Lengthunit=0.9\Lengthunit
\dashlin#1(#2,#3)\mov#1(#2,#3){\sm*}}\mov#1(#2,#3){\sl*
\d**=-.012\Lengthunit\dd*=#2\d**\dt*=#3\d**
\arr*(1,10,4)\arr*(3,8,4)\arr*(4.8,4.2,3)}}}

\def\clap#1{\hbox to 0pt{\hss #1\hss}}

\def\ind(#1,#2)#3{\rlap{\L*=.1\Lengthunit
\xL*=#1\L* \yL*=#2\L* \rx* \the\cos*\xL* \tmp* \the\sin*\yL*
\advance\rx*-\tmp* \ry* \the\cos*\yL* \tmp* \the\sin*\xL*
\advance\ry*\tmp* \kern\rx*\raise\ry*\hbox{\lower2pt\clap{$#3$}}}}

\def\sh*(#1,#2)#3{\rlap{\dm*=\the\n*\d**
\xL*=\xscale\dm* \yL*=\yscale\dm* \xL*=#1\xL* \yL*=#2\yL* \rx*
\the\cos*\xL* \tmp* \the\sin*\yL* \advance\rx*-\tmp* \ry*
\the\cos*\yL* \tmp* \the\sin*\xL* \advance\ry*\tmp*
\kern\rx*\raise\ry*\hbox{#3}}}

\def\calcnum*#1(#2,#3){\a*=1000sp\b*=1000sp\a*=#2\a*\b*=#3\b*
\ifdim\a*<0pt\a*-\a*\fi\ifdim\b*<0pt\b*-\b*\fi
\ifdim\a*>\b*\c*=.96\a*\advance\c*.4\b*
\else\c*=.96\b*\advance\c*.4\a*\fi
\k*\a*\multiply\k*\k*\l*\b*\multiply\l*\l*
\m*\k*\advance\m*\l*\n*\c*\r*\n*\multiply\n*\n*
\dn*\m*\advance\dn*-\n*\divide\dn*2\divide\dn*\r* \advance\r*\dn*
\c*=\the\Nhalfperiods5sp\c*=#1\c*\ifdim\c*<0pt\c*-\c*\fi
\multiply\c*\r*\N*\c*\divide\N*10000}

\def\dashlin#1(#2,#3){\rlap{\calcnum*#1(#2,#3)\relax
\d**=#1\Lengthunit\ifdim\d**<0pt\d**-\d**\fi
\divide\N*2\multiply\N*2\advance\N*\*one
\divide\d**\N*\sm*\n*\*one\sh*(#2,#3){\sl*}\loop
\advance\n*\*one\sh*(#2,#3){\sm*}\advance\n*\*one
\sh*(#2,#3){\sl*}\ifnum\n*<\N*\repeat}}

\def\dashdotlin#1(#2,#3){\rlap{\calcnum*#1(#2,#3)\relax
\d**=#1\Lengthunit\ifdim\d**<0pt\d**-\d**\fi
\divide\N*2\multiply\N*2\advance\N*1\multiply\N*2\relax
\divide\d**\N*\sm*\n*\*two\sh*(#2,#3){\sl*}\loop
\advance\n*\*one\sh*(#2,#3){\kern-1.48pt\lower.5pt\hbox{\rm.}}\relax
\advance\n*\*one\sh*(#2,#3){\sm*}\advance\n*\*two
\sh*(#2,#3){\sl*}\ifnum\n*<\N*\repeat}}

\def\shl*(#1,#2)#3{\kern#1#3\lower#2#3\hbox{\unhcopy\spl*}}

\def\trianglin#1(#2,#3){\rlap{\toks0={#2}\toks1={#3}\calcnum*#1(#2,#3)\relax
\dd*=.57\Lengthunit\dd*=#1\dd*\divide\dd*\N* \divide\dd*\*ths
\multiply\dd*\magnitude
\d**=#1\Lengthunit\ifdim\d**<0pt\d**-\d**\fi
\multiply\N*2\divide\d**\N*\sm*\n*\*one\loop
\shl**{\dd*}\dd*-\dd*\advance\n*2\relax
\ifnum\n*<\N*\repeat\n*\N*\shl**{0pt}}}

\def\wavelin#1(#2,#3){\rlap{\toks0={#2}\toks1={#3}\calcnum*#1(#2,#3)\relax
\dd*=.23\Lengthunit\dd*=#1\dd*\divide\dd*\N* \divide\dd*\*ths
\multiply\dd*\magnitude
\d**=#1\Lengthunit\ifdim\d**<0pt\d**-\d**\fi
\multiply\N*4\divide\d**\N*\sm*\n*\*one\loop
\shl**{\dd*}\dt*=1.3\dd*\advance\n*\*one
\shl**{\dt*}\advance\n*\*one \shl**{\dd*}\advance\n*\*two
\dd*-\dd*\ifnum\n*<\N*\repeat\n*\N*\shl**{0pt}}}

\def\w*lin(#1,#2){\rlap{\toks0={#1}\toks1={#2}\d**=\Lengthunit\dd*=-.12\d**
\divide\dd*\*ths \multiply\dd*\magnitude
\N*8\divide\d**\N*\sm*\n*\*one\loop
\shl**{\dd*}\dt*=1.3\dd*\advance\n*\*one
\shl**{\dt*}\advance\n*\*one \shl**{\dd*}\advance\n*\*one
\shl**{0pt}\dd*-\dd*\advance\n*1\ifnum\n*<\N*\repeat}}

\def\l*arc(#1,#2)[#3][#4]{\rlap{\toks0={#1}\toks1={#2}\d**=\Lengthunit
\dd*=#3.037\d**\dd*=#4\dd*\dt*=#3.049\d**\dt*=#4\dt*\ifdim\d**>10mm\relax
\d**=.25\d**\n*\*one\shl**{-\dd*}\n*\*two\shl**{-\dt*}\n*3\relax
\shl**{-\dd*}\n*4\relax\shl**{0pt}\else
\ifdim\d**>5mm\d**=.5\d**\n*\*one\shl**{-\dt*}\n*\*two
\shl**{0pt}\else\n*\*one\shl**{0pt}\fi\fi}}

\def\d*arc(#1,#2)[#3][#4]{\rlap{\toks0={#1}\toks1={#2}\d**=\Lengthunit
\dd*=#3.037\d**\dd*=#4\dd*\d**=.25\d**\sm*\n*\*one\shl**{-\dd*}\relax
\n*3\relax\sh*(#1,#2){\xL*=\xscale\dd*\yL*=\yscale\dd*
\kern#2\xL*\lower#1\yL*\hbox{\sm*}}\n*4\relax\shl**{0pt}}}

\def\shl**#1{\c*=\the\n*\d**\d*=#1\relax
\a*=\the\toks0\c*\b*=\the\toks1\d*\advance\a*-\b*
\b*=\the\toks1\c*\d*=\the\toks0\d*\advance\b*\d*
\a*=\xscale\a*\b*=\yscale\b* \rx* \the\cos*\a* \tmp* \the\sin*\b*
\advance\rx*-\tmp* \ry* \the\cos*\b* \tmp* \the\sin*\a*
\advance\ry*\tmp* \raise\ry*\rlap{\kern\rx*\unhcopy\spl*}}

\def\wlin*#1(#2,#3)[#4]{\rlap{\toks0={#2}\toks1={#3}\relax
\c*=#1\l*\c*\c*=.01\Lengthunit\m*\c*\divide\l*\m*
\c*=\the\Nhalfperiods5sp\multiply\c*\l*\N*\c*\divide\N*\*ths
\divide\N*2\multiply\N*2\advance\N*\*one
\dd*=.002\Lengthunit\dd*=#4\dd*\multiply\dd*\l*\divide\dd*\N*
\divide\dd*\*ths \multiply\dd*\magnitude
\d**=#1\multiply\N*4\divide\d**\N*\sm*\n*\*one\loop
\shl**{\dd*}\dt*=1.3\dd*\advance\n*\*one
\shl**{\dt*}\advance\n*\*one \shl**{\dd*}\advance\n*\*two
\dd*-\dd*\ifnum\n*<\N*\repeat\n*\N*\shl**{0pt}}}

\def\wavebox#1{\setbox0\hbox{#1}\relax
\a*=\wd0\advance\a*14pt\b*=\ht0\advance\b*\dp0\advance\b*14pt\relax
\hbox{\kern9pt\relax
\rmov*(0pt,\ht0){\rmov*(-7pt,7pt){\wlin*\a*(1,0)[+]\wlin*\b*(0,-1)[-]}}\relax
\rmov*(\wd0,-\dp0){\rmov*(7pt,-7pt){\wlin*\a*(-1,0)[+]\wlin*\b*(0,1)[-]}}\relax
\box0\kern9pt}}

\def\rectangle#1(#2,#3){\relax
\lin#1(#2,0)\lin#1(0,#3)\mov#1(0,#3){\lin#1(#2,0)}\mov#1(#2,0){\lin#1(0,#3)}}

\def\dashrectangle#1(#2,#3){\dashlin#1(#2,0)\dashlin#1(0,#3)\relax
\mov#1(0,#3){\dashlin#1(#2,0)}\mov#1(#2,0){\dashlin#1(0,#3)}}

\def\waverectangle#1(#2,#3){\L*=#1\Lengthunit\a*=#2\L*\b*=#3\L*
\ifdim\a*<0pt\a*-\a*\def\x*{-1}\else\def\x*{1}\fi
\ifdim\b*<0pt\b*-\b*\def\y*{-1}\else\def\y*{1}\fi
\wlin*\a*(\x*,0)[-]\wlin*\b*(0,\y*)[+]\relax
\mov#1(0,#3){\wlin*\a*(\x*,0)[+]}\mov#1(#2,0){\wlin*\b*(0,\y*)[-]}}

\def\calcparab*{\ifnum\n*>\m*\k*\N*\advance\k*-\n*\else\k*\n*\fi
\a*=\the\k* sp\a*=10\a*\b*\dm*\advance\b*-\a*\k*\b*
\a*=\the\*ths\b*\divide\a*\l*\multiply\a*\k*
\divide\a*\l*\k*\*ths\r*\a*\advance\k*-\r*\dt*=\the\k*\L*}

\def\arcto#1(#2,#3)[#4]{\rlap{\toks0={#2}\toks1={#3}\calcnum*#1(#2,#3)\relax
\dm*=135sp\dm*=#1\dm*\d**=#1\Lengthunit\ifdim\dm*<0pt\dm*-\dm*\fi
\multiply\dm*\r*\a*=.3\dm*\a*=#4\a*\ifdim\a*<0pt\a*-\a*\fi
\advance\dm*\a*\N*\dm*\divide\N*10000\relax
\divide\N*2\multiply\N*2\advance\N*\*one
\L*=-.25\d**\L*=#4\L*\divide\d**\N*\divide\L*\*ths
\m*\N*\divide\m*2\dm*=\the\m*5sp\l*\dm*\sm*\n*\*one\loop
\calcparab*\shl**{-\dt*}\advance\n*1\ifnum\n*<\N*\repeat}}

\def\arrarcto#1(#2,#3)[#4]{\L*=#1\Lengthunit\L*=.54\L*
\arcto#1(#2,#3)[#4]\rmov*(#2\L*,#3\L*){\d*=.457\L*\d*=#4\d*\d**-\d*
\rmov*(#3\d**,#2\d*){\arrow.02(#2,#3)}}}

\def\dasharcto#1(#2,#3)[#4]{\rlap{\toks0={#2}\toks1={#3}\relax
\calcnum*#1(#2,#3)\dm*=\the\N*5sp\a*=.3\dm*\a*=#4\a*\ifdim\a*<0pt\a*-\a*\fi
\advance\dm*\a*\N*\dm*
\divide\N*20\multiply\N*2\advance\N*1\d**=#1\Lengthunit
\L*=-.25\d**\L*=#4\L*\divide\d**\N*\divide\L*\*ths
\m*\N*\divide\m*2\dm*=\the\m*5sp\l*\dm*
\sm*\n*\*one\loop\calcparab*
\shl**{-\dt*}\advance\n*1\ifnum\n*>\N*\else\calcparab*
\sh*(#2,#3){\xL*=#3\dt* \yL*=#2\dt* \rx* \the\cos*\xL* \tmp*
\the\sin*\yL* \advance\rx*\tmp* \ry* \the\cos*\yL* \tmp*
\the\sin*\xL* \advance\ry*-\tmp*
\kern\rx*\lower\ry*\hbox{\sm*}}\fi
\advance\n*1\ifnum\n*<\N*\repeat}}

\def\*shl*#1{\c*=\the\n*\d**\advance\c*#1\a**\d*\dt*\advance\d*#1\b**
\a*=\the\toks0\c*\b*=\the\toks1\d*\advance\a*-\b*
\b*=\the\toks1\c*\d*=\the\toks0\d*\advance\b*\d* \rx* \the\cos*\a*
\tmp* \the\sin*\b* \advance\rx*-\tmp* \ry* \the\cos*\b* \tmp*
\the\sin*\a* \advance\ry*\tmp*
\raise\ry*\rlap{\kern\rx*\unhcopy\spl*}}

\def\calcnormal*#1{\b**=10000sp\a**\b**\k*\n*\advance\k*-\m*
\multiply\a**\k*\divide\a**\m*\a**=#1\a**\ifdim\a**<0pt\a**-\a**\fi
\ifdim\a**>\b**\d*=.96\a**\advance\d*.4\b**
\else\d*=.96\b**\advance\d*.4\a**\fi
\d*=.01\d*\r*\d*\divide\a**\r*\divide\b**\r*
\ifnum\k*<0\a**-\a**\fi\d*=#1\d*\ifdim\d*<0pt\b**-\b**\fi
\k*\a**\a**=\the\k*\dd*\k*\b**\b**=\the\k*\dd*}

\def\wavearcto#1(#2,#3)[#4]{\rlap{\toks0={#2}\toks1={#3}\relax
\calcnum*#1(#2,#3)\c*=\the\N*5sp\a*=.4\c*\a*=#4\a*\ifdim\a*<0pt\a*-\a*\fi
\advance\c*\a*\N*\c*\divide\N*20\multiply\N*2\advance\N*-1\multiply\N*4\relax
\d**=#1\Lengthunit\dd*=.012\d** \divide\dd*\*ths
\multiply\dd*\magnitude \ifdim\d**<0pt\d**-\d**\fi\L*=.25\d**
\divide\d**\N*\divide\dd*\N*\L*=#4\L*\divide\L*\*ths
\m*\N*\divide\m*2\dm*=\the\m*0sp\l*\dm*
\sm*\n*\*one\loop\calcnormal*{#4}\calcparab*
\*shl*{1}\advance\n*\*one\calcparab*
\*shl*{1.3}\advance\n*\*one\calcparab*
\*shl*{1}\advance\n*2\dd*-\dd*\ifnum\n*<\N*\repeat\n*\N*\shl**{0pt}}}

\def\triangarcto#1(#2,#3)[#4]{\rlap{\toks0={#2}\toks1={#3}\relax
\calcnum*#1(#2,#3)\c*=\the\N*5sp\a*=.4\c*\a*=#4\a*\ifdim\a*<0pt\a*-\a*\fi
\advance\c*\a*\N*\c*\divide\N*20\multiply\N*2\advance\N*-1\multiply\N*2\relax
\d**=#1\Lengthunit\dd*=.012\d** \divide\dd*\*ths
\multiply\dd*\magnitude \ifdim\d**<0pt\d**-\d**\fi\L*=.25\d**
\divide\d**\N*\divide\dd*\N*\L*=#4\L*\divide\L*\*ths
\m*\N*\divide\m*2\dm*=\the\m*0sp\l*\dm*
\sm*\n*\*one\loop\calcnormal*{#4}\calcparab*
\*shl*{1}\advance\n*2\dd*-\dd*\ifnum\n*<\N*\repeat\n*\N*\shl**{0pt}}}

\def\hr*#1{\L*=\xscale\Lengthunit\ifnum
\angle**=0\clap{\vrule width#1\L* height.1pt}\else
\L*=#1\L*\L*=.5\L*\rmov*(-\L*,0pt){\sm*}\rmov*(\L*,0pt){\sl*}\fi}

\def\shade#1[#2]{\rlap{\Lengthunit=#1\Lengthunit
\special{em:linewidth .001pt}\relax
\mov(0,#2.05){\hr*{.994}}\mov(0,#2.1){\hr*{.980}}\relax
\mov(0,#2.15){\hr*{.953}}\mov(0,#2.2){\hr*{.916}}\relax
\mov(0,#2.25){\hr*{.867}}\mov(0,#2.3){\hr*{.798}}\relax
\mov(0,#2.35){\hr*{.715}}\mov(0,#2.4){\hr*{.603}}\relax
\mov(0,#2.45){\hr*{.435}}\special{em:linewidth \the\linwid*}}}

\def\dshade#1[#2]{\rlap{\special{em:linewidth .001pt}\relax
\Lengthunit=#1\Lengthunit\if#2-\def\t*{+}\else\def\t*{-}\fi
\mov(0,\t*.025){\relax
\mov(0,#2.05){\hr*{.995}}\mov(0,#2.1){\hr*{.988}}\relax
\mov(0,#2.15){\hr*{.969}}\mov(0,#2.2){\hr*{.937}}\relax
\mov(0,#2.25){\hr*{.893}}\mov(0,#2.3){\hr*{.836}}\relax
\mov(0,#2.35){\hr*{.760}}\mov(0,#2.4){\hr*{.662}}\relax
\mov(0,#2.45){\hr*{.531}}\mov(0,#2.5){\hr*{.320}}\relax
\special{em:linewidth \the\linwid*}}}}

\def\vdot{\rlap{\kern-1.9pt\lower1.8pt\hbox{$\scriptstyle\bullet$}}}
\def\vtimes{\rlap{\kern-3pt\lower1.8pt\hbox{$\scriptstyle\times$}}}
\def\vDot{\rlap{\kern-2.3pt\lower2.7pt\hbox{$\bullet$}}}
\def\vTimes{\rlap{\kern-3.6pt\lower2.4pt\hbox{$\times$}}}

\def\arc(#1)[#2,#3]{{\k*=#2\l*=#3\m*=\l*
\advance\m*-6\ifnum\k*>\l*\relax\else
{\rotate(#2)\mov(#1,0){\sm*}}\loop
\ifnum\k*<\m*\advance\k*5{\rotate(\k*)\mov(#1,0){\sl*}}\repeat
{\rotate(#3)\mov(#1,0){\sl*}}\fi}}

\def\dasharc(#1)[#2,#3]{{\k**=#2\n*=#3\advance\n*-1\advance\n*-\k**
\L*=1000sp\L*#1\L* \multiply\L*\n* \multiply\L*\Nhalfperiods
\divide\L*57\N*\L* \divide\N*2000\ifnum\N*=0\N*1\fi \r*\n*
\divide\r*\N* \ifnum\r*<2\r*2\fi \m**\r* \divide\m**2 \l**\r*
\advance\l**-\m** \N*\n* \divide\N*\r* \k**\r* \multiply\k**\N*
\dn*\n* \advance\dn*-\k** \divide\dn*2\advance\dn*\*one \r*\l**
\divide\r*2\advance\dn*\r* \advance\N*-2\k**#2\relax
\ifnum\l**<6{\rotate(#2)\mov(#1,0){\sm*}}\advance\k**\dn*
{\rotate(\k**)\mov(#1,0){\sl*}}\advance\k**\m**
{\rotate(\k**)\mov(#1,0){\sm*}}\loop
\advance\k**\l**{\rotate(\k**)\mov(#1,0){\sl*}}\advance\k**\m**
{\rotate(\k**)\mov(#1,0){\sm*}}\advance\N*-1\ifnum\N*>0\repeat
{\rotate(#3)\mov(#1,0){\sl*}}\else\advance\k**\dn*
\arc(#1)[#2,\k**]\loop\advance\k**\m** \r*\k** \advance\k**\l**
{\arc(#1)[\r*,\k**]}\relax \advance\N*-1\ifnum\N*>0\repeat
\advance\k**\m**\arc(#1)[\k**,#3]\fi}}

\def\triangarc#1(#2)[#3,#4]{{\k**=#3\n*=#4\advance\n*-\k**
\L*=1000sp\L*#2\L* \multiply\L*\n* \multiply\L*\Nhalfperiods
\divide\L*57\N*\L* \divide\N*1000\ifnum\N*=0\N*1\fi
\d**=#2\Lengthunit \d*\d** \divide\d*57\multiply\d*\n* \r*\n*
\divide\r*\N* \ifnum\r*<2\r*2\fi \m**\r* \divide\m**2 \l**\r*
\advance\l**-\m** \N*\n* \divide\N*\r* \dt*\d* \divide\dt*\N*
\dt*.5\dt* \dt*#1\dt* \divide\dt*1000\multiply\dt*\magnitude
\k**\r* \multiply\k**\N* \dn*\n* \advance\dn*-\k**
\divide\dn*2\relax \r*\l** \divide\r*2\advance\dn*\r*
\advance\N*-1\k**#3\relax
{\rotate(#3)\mov(#2,0){\sm*}}\advance\k**\dn*
{\rotate(\k**)\mov(#2,0){\sl*}}\advance\k**-\m**\advance\l**\m**\loop\dt*-\dt*
\d*\d** \advance\d*\dt*
\advance\k**\l**{\rotate(\k**)\rmov*(\d*,0pt){\sl*}}%
\advance\N*-1\ifnum\N*>0\repeat\advance\k**\m**
{\rotate(\k**)\mov(#2,0){\sl*}}{\rotate(#4)\mov(#2,0){\sl*}}}}

\def\wavearc#1(#2)[#3,#4]{{\k**=#3\n*=#4\advance\n*-\k**
\L*=4000sp\L*#2\L* \multiply\L*\n* \multiply\L*\Nhalfperiods
\divide\L*57\N*\L* \divide\N*1000\ifnum\N*=0\N*1\fi
\d**=#2\Lengthunit \d*\d** \divide\d*57\multiply\d*\n* \r*\n*
\divide\r*\N* \ifnum\r*=0\r*1\fi \m**\r* \divide\m**2 \l**\r*
\advance\l**-\m** \N*\n* \divide\N*\r* \dt*\d* \divide\dt*\N*
\dt*.7\dt* \dt*#1\dt* \divide\dt*1000\multiply\dt*\magnitude
\k**\r* \multiply\k**\N* \dn*\n* \advance\dn*-\k**
\divide\dn*2\relax \divide\N*4\advance\N*-1\k**#3\relax
{\rotate(#3)\mov(#2,0){\sm*}}\advance\k**\dn*
{\rotate(\k**)\mov(#2,0){\sl*}}\advance\k**-\m**\advance\l**\m**\loop\dt*-\dt*
\d*\d** \advance\d*\dt* \dd*\d** \advance\dd*1.3\dt*
\advance\k**\r*{\rotate(\k**)\rmov*(\d*,0pt){\sl*}}\relax
\advance\k**\r*{\rotate(\k**)\rmov*(\dd*,0pt){\sl*}}\relax
\advance\k**\r*{\rotate(\k**)\rmov*(\d*,0pt){\sl*}}\relax
\advance\k**\r* \advance\N*-1\ifnum\N*>0\repeat\advance\k**\m**
{\rotate(\k**)\mov(#2,0){\sl*}}{\rotate(#4)\mov(#2,0){\sl*}}}}

\def\gmov*#1(#2,#3)#4{\rlap{\L*=#1\Lengthunit
\xL*=#2\L* \yL*=#3\L* \rx* \gcos*\xL* \tmp* \gsin*\yL*
\advance\rx*-\tmp* \ry* \gcos*\yL* \tmp* \gsin*\xL*
\advance\ry*\tmp* \rx*=\xscale\rx* \ry*=\yscale\ry* \xL*
\the\cos*\rx* \tmp* \the\sin*\ry* \advance\xL*-\tmp* \yL*
\the\cos*\ry* \tmp* \the\sin*\rx* \advance\yL*\tmp*
\kern\xL*\raise\yL*\hbox{#4}}}

\def\rgmov*(#1,#2)#3{\rlap{\xL*#1\yL*#2\relax
\rx* \gcos*\xL* \tmp* \gsin*\yL* \advance\rx*-\tmp* \ry*
\gcos*\yL* \tmp* \gsin*\xL* \advance\ry*\tmp* \rx*=\xscale\rx*
\ry*=\yscale\ry* \xL* \the\cos*\rx* \tmp* \the\sin*\ry*
\advance\xL*-\tmp* \yL* \the\cos*\ry* \tmp* \the\sin*\rx*
\advance\yL*\tmp* \kern\xL*\raise\yL*\hbox{#3}}}

\def\Earc(#1)[#2,#3][#4,#5]{{\k*=#2\l*=#3\m*=\l*
\advance\m*-6\ifnum\k*>\l*\relax\else\def\xscale{#4}\def\yscale{#5}\relax
{\angle**0\rotate(#2)}\gmov*(#1,0){\sm*}\loop
\ifnum\k*<\m*\advance\k*5\relax
{\angle**0\rotate(\k*)}\gmov*(#1,0){\sl*}\repeat
{\angle**0\rotate(#3)}\gmov*(#1,0){\sl*}\relax
\def\xscale{1}\def\yscale{1}\fi}}

\def\dashEarc(#1)[#2,#3][#4,#5]{{\k**=#2\n*=#3\advance\n*-1\advance\n*-\k**
\L*=1000sp\L*#1\L* \multiply\L*\n* \multiply\L*\Nhalfperiods
\divide\L*57\N*\L* \divide\N*2000\ifnum\N*=0\N*1\fi \r*\n*
\divide\r*\N* \ifnum\r*<2\r*2\fi \m**\r* \divide\m**2 \l**\r*
\advance\l**-\m** \N*\n* \divide\N*\r* \k**\r*\multiply\k**\N*
\dn*\n* \advance\dn*-\k** \divide\dn*2\advance\dn*\*one \r*\l**
\divide\r*2\advance\dn*\r* \advance\N*-2\k**#2\relax
\ifnum\l**<6\def\xscale{#4}\def\yscale{#5}\relax
{\angle**0\rotate(#2)}\gmov*(#1,0){\sm*}\advance\k**\dn*
{\angle**0\rotate(\k**)}\gmov*(#1,0){\sl*}\advance\k**\m**
{\angle**0\rotate(\k**)}\gmov*(#1,0){\sm*}\loop
\advance\k**\l**{\angle**0\rotate(\k**)}\gmov*(#1,0){\sl*}\advance\k**\m**
{\angle**0\rotate(\k**)}\gmov*(#1,0){\sm*}\advance\N*-1\ifnum\N*>0\repeat
{\angle**0\rotate(#3)}\gmov*(#1,0){\sl*}\def\xscale{1}\def\yscale{1}\else
\advance\k**\dn* \Earc(#1)[#2,\k**][#4,#5]\loop\advance\k**\m**
\r*\k** \advance\k**\l** {\Earc(#1)[\r*,\k**][#4,#5]}\relax
\advance\N*-1\ifnum\N*>0\repeat
\advance\k**\m**\Earc(#1)[\k**,#3][#4,#5]\fi}}

\def\triangEarc#1(#2)[#3,#4][#5,#6]{{\k**=#3\n*=#4\advance\n*-\k**
\L*=1000sp\L*#2\L* \multiply\L*\n* \multiply\L*\Nhalfperiods
\divide\L*57\N*\L* \divide\N*1000\ifnum\N*=0\N*1\fi
\d**=#2\Lengthunit \d*\d** \divide\d*57\multiply\d*\n* \r*\n*
\divide\r*\N* \ifnum\r*<2\r*2\fi \m**\r* \divide\m**2 \l**\r*
\advance\l**-\m** \N*\n* \divide\N*\r* \dt*\d* \divide\dt*\N*
\dt*.5\dt* \dt*#1\dt* \divide\dt*1000\multiply\dt*\magnitude
\k**\r* \multiply\k**\N* \dn*\n* \advance\dn*-\k**
\divide\dn*2\relax \r*\l** \divide\r*2\advance\dn*\r*
\advance\N*-1\k**#3\relax
\def\xscale{#5}\def\yscale{#6}\relax
{\angle**0\rotate(#3)}\gmov*(#2,0){\sm*}\advance\k**\dn*
{\angle**0\rotate(\k**)}\gmov*(#2,0){\sl*}\advance\k**-\m**
\advance\l**\m**\loop\dt*-\dt* \d*\d** \advance\d*\dt*
\advance\k**\l**{\angle**0\rotate(\k**)}\rgmov*(\d*,0pt){\sl*}\relax
\advance\N*-1\ifnum\N*>0\repeat\advance\k**\m**
{\angle**0\rotate(\k**)}\gmov*(#2,0){\sl*}\relax
{\angle**0\rotate(#4)}\gmov*(#2,0){\sl*}\def\xscale{1}\def\yscale{1}}}

\def\waveEarc#1(#2)[#3,#4][#5,#6]{{\k**=#3\n*=#4\advance\n*-\k**
\L*=4000sp\L*#2\L* \multiply\L*\n* \multiply\L*\Nhalfperiods
\divide\L*57\N*\L* \divide\N*1000\ifnum\N*=0\N*1\fi
\d**=#2\Lengthunit \d*\d** \divide\d*57\multiply\d*\n* \r*\n*
\divide\r*\N* \ifnum\r*=0\r*1\fi \m**\r* \divide\m**2 \l**\r*
\advance\l**-\m** \N*\n* \divide\N*\r* \dt*\d* \divide\dt*\N*
\dt*.7\dt* \dt*#1\dt* \divide\dt*1000\multiply\dt*\magnitude
\k**\r* \multiply\k**\N* \dn*\n* \advance\dn*-\k**
\divide\dn*2\relax
\divide\N*4\advance\N*-1\k**#3\def\xscale{#5}\def\yscale{#6}\relax
{\angle**0\rotate(#3)}\gmov*(#2,0){\sm*}\advance\k**\dn*
{\angle**0\rotate(\k**)}\gmov*(#2,0){\sl*}\advance\k**-\m**
\advance\l**\m**\loop\dt*-\dt* \d*\d** \advance\d*\dt* \dd*\d**
\advance\dd*1.3\dt*
\advance\k**\r*{\angle**0\rotate(\k**)}\rgmov*(\d*,0pt){\sl*}\relax
\advance\k**\r*{\angle**0\rotate(\k**)}\rgmov*(\dd*,0pt){\sl*}\relax
\advance\k**\r*{\angle**0\rotate(\k**)}\rgmov*(\d*,0pt){\sl*}\relax
\advance\k**\r* \advance\N*-1\ifnum\N*>0\repeat\advance\k**\m**
{\angle**0\rotate(\k**)}\gmov*(#2,0){\sl*}\relax
{\angle**0\rotate(#4)}\gmov*(#2,0){\sl*}\def\xscale{1}\def\yscale{1}}}

\newcount\CatcodeOfAtSign
\CatcodeOfAtSign=\the\catcode`\@ \catcode`\@=11
\def\@arc#1[#2][#3]{\rlap{\Lengthunit=#1\Lengthunit
\sm*\l*arc(#2.1914,#3.0381)[#2][#3]\relax
\mov(#2.1914,#3.0381){\l*arc(#2.1622,#3.1084)[#2][#3]}\relax
\mov(#2.3536,#3.1465){\l*arc(#2.1084,#3.1622)[#2][#3]}\relax
\mov(#2.4619,#3.3086){\l*arc(#2.0381,#3.1914)[#2][#3]}}}

\def\dash@arc#1[#2][#3]{\rlap{\Lengthunit=#1\Lengthunit
\d*arc(#2.1914,#3.0381)[#2][#3]\relax
\mov(#2.1914,#3.0381){\d*arc(#2.1622,#3.1084)[#2][#3]}\relax
\mov(#2.3536,#3.1465){\d*arc(#2.1084,#3.1622)[#2][#3]}\relax
\mov(#2.4619,#3.3086){\d*arc(#2.0381,#3.1914)[#2][#3]}}}

\def\wave@arc#1[#2][#3]{\rlap{\Lengthunit=#1\Lengthunit
\w*lin(#2.1914,#3.0381)\relax
\mov(#2.1914,#3.0381){\w*lin(#2.1622,#3.1084)}\relax
\mov(#2.3536,#3.1465){\w*lin(#2.1084,#3.1622)}\relax
\mov(#2.4619,#3.3086){\w*lin(#2.0381,#3.1914)}}}

\def\bezier#1(#2,#3)(#4,#5)(#6,#7){\N*#1\l*\N* \advance\l*\*one
\d* #4\Lengthunit \advance\d* -#2\Lengthunit \multiply\d* \*two
\b* #6\Lengthunit \advance\b* -#2\Lengthunit \advance\b*-\d*
\divide\b*\N* \d** #5\Lengthunit \advance\d** -#3\Lengthunit
\multiply\d** \*two \b** #7\Lengthunit \advance\b** -#3\Lengthunit
\advance\b** -\d** \divide\b**\N*
\mov(#2,#3){\sm*{\loop\ifnum\m*<\l* \a*\m*\b* \advance\a*\d*
\divide\a*\N* \multiply\a*\m* \a**\m*\b** \advance\a**\d**
\divide\a**\N* \multiply\a**\m*
\rmov*(\a*,\a**){\unhcopy\spl*}\advance\m*\*one\repeat}}}

\catcode`\*=12

\newcount\n@ast
\def\n@ast@#1{\n@ast0\relax\get@ast@#1\end}
\def\get@ast@#1{\ifx#1\end\let\next\relax\else
\ifx#1*\advance\n@ast1\fi\let\next\get@ast@\fi\next}

\newif\if@up \newif\if@dwn
\def\up@down@#1{\@upfalse\@dwnfalse
\if#1u\@uptrue\fi\if#1U\@uptrue\fi\if#1+\@uptrue\fi
\if#1d\@dwntrue\fi\if#1D\@dwntrue\fi\if#1-\@dwntrue\fi}

\def\halfcirc#1(#2)[#3]{{\Lengthunit=#2\Lengthunit\up@down@{#3}\relax
\if@up\mov(0,.5){\@arc[-][-]\@arc[+][-]}\fi
\if@dwn\mov(0,-.5){\@arc[-][+]\@arc[+][+]}\fi
\def\lft{\mov(0,.5){\@arc[-][-]}\mov(0,-.5){\@arc[-][+]}}\relax
\def\rght{\mov(0,.5){\@arc[+][-]}\mov(0,-.5){\@arc[+][+]}}\relax
\if#3l\lft\fi\if#3L\lft\fi\if#3r\rght\fi\if#3R\rght\fi
\n@ast@{#1}\relax
\ifnum\n@ast>0\if@up\shade[+]\fi\if@dwn\shade[-]\fi\fi
\ifnum\n@ast>1\if@up\dshade[+]\fi\if@dwn\dshade[-]\fi\fi}}

\def\halfdashcirc(#1)[#2]{{\Lengthunit=#1\Lengthunit\up@down@{#2}\relax
\if@up\mov(0,.5){\dash@arc[-][-]\dash@arc[+][-]}\fi
\if@dwn\mov(0,-.5){\dash@arc[-][+]\dash@arc[+][+]}\fi
\def\lft{\mov(0,.5){\dash@arc[-][-]}\mov(0,-.5){\dash@arc[-][+]}}\relax
\def\rght{\mov(0,.5){\dash@arc[+][-]}\mov(0,-.5){\dash@arc[+][+]}}\relax
\if#2l\lft\fi\if#2L\lft\fi\if#2r\rght\fi\if#2R\rght\fi}}

\def\halfwavecirc(#1)[#2]{{\Lengthunit=#1\Lengthunit\up@down@{#2}\relax
\if@up\mov(0,.5){\wave@arc[-][-]\wave@arc[+][-]}\fi
\if@dwn\mov(0,-.5){\wave@arc[-][+]\wave@arc[+][+]}\fi
\def\lft{\mov(0,.5){\wave@arc[-][-]}\mov(0,-.5){\wave@arc[-][+]}}\relax
\def\rght{\mov(0,.5){\wave@arc[+][-]}\mov(0,-.5){\wave@arc[+][+]}}\relax
\if#2l\lft\fi\if#2L\lft\fi\if#2r\rght\fi\if#2R\rght\fi}}

\catcode`\*=11

\def\Circle#1(#2){\halfcirc#1(#2)[u]\halfcirc#1(#2)[d]\n@ast@{#1}\relax
\ifnum\n@ast>0\L*=\xscale\Lengthunit \ifnum\angle**=0\clap{\vrule
width#2\L* height.1pt}\else
\L*=#2\L*\L*=.5\L*\special{em:linewidth .001pt}\relax
\rmov*(-\L*,0pt){\sm*}\rmov*(\L*,0pt){\sl*}\relax
\special{em:linewidth \the\linwid*}\fi\fi}

\catcode`\*=12

\def\wavecirc(#1){\halfwavecirc(#1)[u]\halfwavecirc(#1)[d]}

\def\dashcirc(#1){\halfdashcirc(#1)[u]\halfdashcirc(#1)[d]}

\def\xscale{1}
\def\yscale{1}

\def\Ellipse#1(#2)[#3,#4]{\def\xscale{#3}\def\yscale{#4}\relax
\Circle#1(#2)\def\xscale{1}\def\yscale{1}}

\def\dashEllipse(#1)[#2,#3]{\def\xscale{#2}\def\yscale{#3}\relax
\dashcirc(#1)\def\xscale{1}\def\yscale{1}}

\def\waveEllipse(#1)[#2,#3]{\def\xscale{#2}\def\yscale{#3}\relax
\wavecirc(#1)\def\xscale{1}\def\yscale{1}}

\def\halfEllipse#1(#2)[#3][#4,#5]{\def\xscale{#4}\def\yscale{#5}\relax
\halfcirc#1(#2)[#3]\def\xscale{1}\def\yscale{1}}

\def\halfdashEllipse(#1)[#2][#3,#4]{\def\xscale{#3}\def\yscale{#4}\relax
\halfdashcirc(#1)[#2]\def\xscale{1}\def\yscale{1}}

\def\halfwaveEllipse(#1)[#2][#3,#4]{\def\xscale{#3}\def\yscale{#4}\relax
\halfwavecirc(#1)[#2]\def\xscale{1}\def\yscale{1}}

\catcode`\@=\the\CatcodeOfAtSign

\abstract{We develop a systematic approach to construct
the one-loop ${\cal N}=4$ SYM effective action
depending on both ${\cal N}=2$ vector multiplet and
hypermultiplet background fields. Beginning with the formulation
of ${\cal N}=4$ SYM theory in terms of ${\cal N}=2$ harmonic
superfields, we construct the one-loop effective action using the
covariant ${\cal N}=2$ harmonic supergraphs and calculate it
in ${\cal N}=2$ harmonic superfield form for
constant Abelian strength $F_{mn}$ and corresponding constant
hypermultiplet fields. The hypermultiplet-dependent effective
action is derived and given by integral over the analytic subspace of
harmonic superspace. We show that each term in the Schwinger-De Witt
expansion of the low-energy effective action is written as
integral over full ${\cal N}=2$ superspace.}


\begin{document}

\newcommand{\be}{\begin{equation}}
\newcommand{\ee}{\end{equation}}
\newcommand{\bea}{\begin{eqnarray}}
\newcommand{\eea}{\end{eqnarray}}


\section{Introduction}


${\cal N}=4$ supersymmetric Yang-Mills (SYM) theory attracts a lot of
attention due to its unique properties in the quantum domain such
as finiteness and superconformal invariance and the remarkable
links with string/brane theory (see e.g. the  \cite{T}, \cite{HF}
for review). Discovery of AdS/CFT correspondence stimulated  a new
significant interest to study the various aspects of ${\cal N}=4$
SYM theory.  AdS/CFT correspondence \cite{M} states a duality
between IIB superstring theory compactified on $AdS_5\times S^5$
and four-dimensional ${\cal N}=4$ SYM theory in 't Hooft limit. It
turnes out, that the low-energy properties of bulk string theory are
related to ${\cal N}=4$ supersymmetric gauge quantum field
theory. To be more precise, in the limit under consideration, the
bulk theory reduces to the higher dimensional classical
supergravity encoded by correlation  functions of gauge invariant
composite operators in $D=4$, ${\cal N}=4$ SYM theory. Another
implementation of ${\cal N}=4$ SYM model to string theory is related
to the conjecture that $D3$-brane interactions in static limit are
completely described in terms of low-energy $D=4$, ${\cal N}=4$
SYM effective action on the Coulomb branch \cite{chep}, \cite{bkt},
\cite{bpt}. All this allows us to treat ${\cal N}=4$ SYM quantum
field model as an element of superstring theory.

Formulation of ${\cal N}=4$ SYM theory possessing manifest
off-shell ${\cal N}=4$ supersymmetry is unknown so far. Superfield
description of  the $D=4, {\cal N}=4$ SYM theory is realized in
terms of  scalar superfield $W_{A B}=-W_{B A}, A,B=1, ...,4$ which
transforms under the six-dimensional representation of the $SU(4)$
internal symmetry group and which is real:
$\bar{W}^{AB}=\frac{1}{2}\epsilon^{ABCD}W_{CD}$. This superfield
obeys the constrains $$ {\cal D}_{\alpha A}W_{BC}={\cal D}_{\alpha
[A} W_{BC]}, \quad \bar{\cal D}^A_{\dot\alpha} W_{BC}=-\frac{2}{3}
\delta^A_{[B} \bar{\cal D}^E_{\dot\alpha}W_{C]E}. $$ These
constraints imply that the component content of the superfield
$W_{AB}$ corresponds to on-shell vector multiplet and includes 6
real scalar fields, 4 Majorana spinor fields and 1 vector
field\footnote{The same multiplet can be described off shell in
${\cal N}=3$ harmonic superspace \cite{gikos1}, \cite{gios}.
Quantum aspects of ${\cal N}=3$ SYM theory have been discussed in
\cite{del}. Structure of effective action of such a theory was
studied in \cite{iz}, \cite{bisz}.}.

From the ${\cal N}=2$ supersymmetric point of view, the ${\cal
N}=4$ vector multiplet consists of the ${\cal N}=2$ vector
multiplet and the hypermultiplet \cite{gios}. Therefore the ${\cal
N}=4$ SYM action can be treated as some special ${\cal N}=2$
supersymmetric theory, with action containing the action of ${\cal
N}=2$ SYM theory plus the action of a hypermultiplet in adjoint
representation coupled to ${\cal N}=2$ vector multiplet. In
addition, this model possesses the hidden ${\cal N}=2$ symmetry
and as a result it actually is ${\cal N}=4$ supersymmetric. Such a
theory is naturally formulated in ${\cal N}=2$ harmonic superspace
\cite{gikos}, \cite{gios}. This formulation simplifies a quantum
consideration due to manifest ${\cal N}=2$ supersymmetry. Analysis
of effective action is more simplified with help of harmonic
superfield background field method \cite{backgr}, \cite{bbiko}.
Straightforward calculation of higher-loop contributions to
effective action in closed form is a very complicated technical
problem. We point out that recently an essential progress in
development of multi-loop calculations techniques in the harmonic
superspace was achieved  for ${\cal N}=2$ vector multiplet
background  \cite{2loop}. Therefore, a study of general structure
of possible higher order corrections to the effective action would
be very useful (see e.g. \cite{howe}).

In this paper the ${\cal N}=4$ SYM theory is treated as some
special case of generic ${\cal N}=2$ gauge models. All such models
are formulated in ${\cal N}=2$ harmonic superspace in manifest
${\cal N}=2$ supersymmetric form. ${\cal N}=4$ SYM theory is
selected up to an extra hidden on-shell ${\cal N}=2$
supersymmetry. Formulation in terms of ${\cal N}=2$ superspace
allows us to use the known classification of ground state
structure in ${\cal N}=2$ gauge models (see e.g. \cite{ARG}).
According to such a classification, the branch of the theory
possessing the non-zero vacuum expectation values both for scalar
from N=2 vector multiplet and for hypermultiplet scalar is called
the mixed branch. This just corresponds to the problem under
consideration. It is clear, to keep the hidden ${\cal N}=2$
supersymmetry of the theory in ground state we have to preserve
non-vanishing expectation values not only for ${\cal N}=2$ vector
multiplet but also for hypermultiplet. The ${\cal N}=2$ hidden
supersymmetry transformation of values in ground state and their
applications were given in \cite{bbp}\footnote{We would like to
pay attention that there is no generally accepted terminology in
literature concerning the ground state of ${\cal N}=4$ SYM theory.
One possibilities is to use the terminology from ${\cal N}=2$
gauge theories since ${\cal N}=4$ SYM is a partial case of these
theories. There we follow namely such a possibility}.

At present, it is well known that the exact low-energy quantum
dynamics of ${\cal N}=4$ SYM theory in ${\cal N}=2$ vector
multiplet sector for $SU(N)$ gauge group spontaneously broken down
to its maximal torus $U(1)^{N-1}$  is described by the
non-holomorphic effective potential ${\cal H}({\cal W},{\bar{\cal
W}})$, depending on ${\cal N}=2$ strengths ${\cal W}, {\bar{\cal
W}}$ \cite{6a}, \cite{bbku}, \cite{bbiko}. We emphasize that the
structure of  ${\cal H}({\cal W},{\bar{\cal W}})$ is so unique
that it can be obtained entirely on the symmetry grounds of scale
independence and R-invariance up to a numerical factor \cite{6a},
\cite{nonr}.  Moreover, the potential ${\cal H}({\cal
W},{\bar{\cal W}})$ gets neither perturbative quantum corrections
beyond one-loop nor instanton corrections \cite{6a}, (see also
discussion of non-holomorphic potential in ${\cal N}=2$ SYM
theories \cite{nonr}, \cite{dwgr} and structure of next-to-leading
two-loop contributions to effective action \cite{Bec}, \cite{bpt},
\cite{2loop}). All these properties are very important for
understanding of the low-energy quantum dynamics in ${\cal N}=4$
SYM theory and its applications. In particular, the effective
potential ${\cal H}({\cal W},{\bar{\cal W}})$ describes  the
leading terms in the interaction between parallel D3-branes in
superstring theory.

To clarify the restrictions on the effective action, stipulated by
${\cal N}=4$ supersymmetry, to describe complete structure of the
effective action and to gain a deeper understanding of the ${\cal
N}=4$ SYM/supergravity correspondence, we have to find an effective
action not only in ${\cal N}=2$ vector multiplet sector but its
dependence on all the fields of ${\cal N}=4$ vector multiplet. The
problem of the ${\cal N}=4$ SYM effective action on the so called
mixed branch remained unstudied for a long time. Recently, the
complete exact low-energy effective potential ${\cal L}_q (X)$
($X=-\frac{q^{ia}q_{ia}}{{\cal W}\bar{\cal W}}$) containing the
dependence both on ${\cal N}=2$ gauge superfields and
hypermultiplets has been obtained \cite{1}. It has been shown
that the algebraic restrictions imposed by the hidden ${\cal N}=2$
supersymmetry on a structure of the low-energy effective action in
${\cal N}=2$ harmonic superspace approach turns out to be so strong
that they allow us  to restore the dependence of the low-energy
effective action on the hypermultiplets on basis of the known
non-holomorphic effective potential ${\cal H}({\cal W}, \bar{\cal
W})$. As a result, the additional hypermultiplet-dependent
contribution to low-energy effective action containing the
on-shell ${\cal W}, {\bar{\cal W}}$ and the hypermultiplet
$q^{ia}$ superfields has been obtained. The leading low-energy
effective Lagrangian ${\cal L}_q (X)$ was found in Ref. \cite{1}
on a purely algebraic ground. Then, it was shown in paper
\cite{1b}, that the  effective Lagrangian ${\cal L}_q (X)$ can be
derived using the harmonic supergraph techniques and harmonic
superspace background field method. Later, the structure of
the one-loop effective action beyond leading approximation has
been found in \cite{bbp} on the base of formulation of ${\cal
N}=4$ SYM theory in terms of ${\cal N}=1$ superfields and
exploring the derivative expansion techniques in ${\cal N}=1$
superspace \cite{pb}. Although such a formulation preserves a fewer
number of manifest supersymmetries than in the harmonic superspace
approach, the supersymmetric $R_\xi$-gauges and a special
prescription on restoration of the ${\cal N}=2$ supersymmetric
form make possible to construct the effective action including
dependence on arbitrary powers of the Abelian strength $F_{mn}$ and
special $R$-symmetry invariant combination of the constant
component fields $\phi, f^{ia}$ of hypermultiplet.

The present paper is devoted to analysis of hypermultiplet
dependence of ${\cal N}=4$ SYM low-energy effective action in
${\cal N}=2$ harmonic superspace. We compute the low-energy
effective action for the space-time independent ${\cal N}=2$
vector multiplet and  hypermultiplet background under the
followig assumptions: (i)
${\cal W}|_{\theta=0}=\mbox{const}, D^{\pm}_{\alpha}{\cal
W}|_{\theta=0}=\mbox{const}, D^-_\alpha D^+_\beta {\cal
W}|_{\theta=0}=\mbox{const}, 
$ (ii) the background hypermultiplet is on-shell, $q^{\pm a} =
q^{ia}u^{\pm}_{i}$ where $q^{ia}$ does not depend on harmonics \cite{1}
and is considered constant. This means that, in this paper, we
treat the effective action in the hypermultiplet sector as a series in spinor derivatives
of $q^{+a}$ and study the simplest approximation corresponding to
contributions
to the effective action which do not depend on any derivatives of $q^{+a}$.
Comparing with paper \cite{1b}, we consider the effective action
beyond the leading low-energy approximation in ${\cal N}=2$ vector multiplet sector
and take into account
all powers of the Abelian strength $F_{mn}$. On the other hand,
comparing with paper \cite{bbp}, we work completely in terms of
${\cal N}=2$ harmonic superspace at all steps of consideration and
justify the special heuristic prescription concerning a
restoration of manifest ${\cal N}=2$ supersymmetric form of
effective action, which has been used in \cite{bbp}. As a result
we obtain a proper-time representation of low-energy effective
action written as integral over the analytic subspace of harmonic
superspace.

This paper is organized as follows. In Section 2, we discuss the
construction of ${\cal N}=4$ SYM model in ${\cal N}=2$ harmonic
superspace and structure of perturbation theory in this  model.
Section 3 is devoted to a generic procedure of finding  the one-loop
effective action in the hypermultiplet sector. In Section 4, we show
how one can sum up an infinite series of covariant harmonic
supergraphs with arbitrary number of hypermultiplet legs on
non-trivial ${\cal N}=2$ vector multiplet background to get a form
of the effective action which can be studied on the base of
proper-time method. Section 5 is devoted to calculating the
one-loop effective action with help of symbol operator techniques
we develop in ${\cal N}=2$ harmonic superspace. In Section 6, we
derive the final result for the one-loop effective action in the form
of an integral over the analytic subspace of harmonic superspace. We also
show that this result leads to spinor covariant derivative
expansion of the effective action and find two first terms of this
expansion in explicit form. We demonstrate that each term of the
expansion can be rewritten as integral over full ${\cal N}$=2
superspace. In Section 7, we summarize our main results.


\section{${\cal N}=4$ SYM theory in ${\cal N}=2$ harmonic superspace}


The harmonic superspace is an universal construction for
formulating the arbitrary ${\cal N}=2$ supersymmetric theories in
a manifestly ${\cal N}=2$ supersymmetric way. ${\cal N}=2$
harmonic superspace was introduced in Ref. \cite{gikos} to
construct off-shell ${\cal N}=2$ multiplets and develop a
superfield formulation of ${\cal N}=2$ models in terms of
unconstrained superfields. This approach extends the conventional
${\cal N}=2$ superspace $z=(x^{m}, \theta^\alpha_i,
\bar{\theta}^{\dot\alpha i})$ by the two-sphere $S^2 =SU(2)/U(1)$
parameterized by harmonics $u^{+i}$ and their conjugate $u^-_i$.
Throughout this paper we follow the conventions of the book
\cite{gios}. Harmonic superspace with coordinates $z, u$ contains the analytic
subspace parameterized by the variables $\zeta=(x_A^m,
\theta^{+\alpha}, \bar{\theta}^+_{\dot\alpha}, u^+_i, u^-_i ),$
where the analytic basis is defined by $$x_A^m = x^m - 2i
\theta^{(i}\sigma^m \bar{\theta}^{j)} u^+_i u^-_j,
\theta^{\pm}_\alpha = u^{\pm}_i \theta^i_\alpha,
\bar{\theta}^{\pm}_{\dot\alpha} =
u^{\pm}_i\bar{\theta}^i_{\dot\alpha}.$$ The spinor covariant
derivatives in the central and the analytic bases (see discussion
of these bases in \cite{gios}) are related by ${D}^{\pm}_\alpha =
u^{\pm}_i {D}^i_\alpha, \bar{D}^{\pm}_{\dot\alpha} = u^{\pm}_i
{D}^i_{\dot\alpha}.$ A very important feature is that the
operators $D^+_\alpha, \bar{D}^+_{\dot\alpha}$ strictly
anticommuting. A covariantly analytic superfield
$\Phi^{(p)}(z,u)$\footnote{Here the superscript $p$ refers to the
harmonic $U(1)$ charge, $D^0\Phi^{(p)}=p \Phi^{(p)}$} is defined
to be annihilated by these operators, $D^+_\alpha
\Phi^{(p)}=\bar{D}^+_{\dot\alpha} \Phi^{(p)}=0$.

The two basic ${\cal N}=2$ ingredients of the ${\cal N}=4$ SYM
theory are a hypermultiplet and a ${\cal N}=2$ vector multiplet.
In harmonic superspace  hypermultiplets are described by an
analytic superfield of $U(1)$ charge equal to $+1$, $$D^+_\alpha
q^+ = \bar{D}^+_{\dot\alpha} q^+ = 0 \Rightarrow q^+ = q^+ (x_A,
\theta^+, \bar{\theta}^+, u).$$ In the on-shell case the
hyper\-multi\-plet sa\-tis\-fies the con\-di\-tion of har\-mo\-nic
ana\-ly\-ti\-city\\ $D^{++} q^+ = 0$ and forms the so called an
ultrashort superfield. Here $$D^{++} = u^{+i}
\frac{\partial}{\partial u^{-i}} -2i (\theta^+
\sigma^m\bar{\theta}^+) \frac{\partial}{\partial x_A^m}$$ is the
covariant harmonic derivative in the analytic basis, which also
plays the role of raising operator in the $su(2)$ algebra:
$[D^{++}, D^{--}] = D^0,$ $[D^0, D^{\pm \pm}] = \pm  2
D^{\pm\pm},$ $[D^{++}, D^-] = D^+$.  The classical action of the
hypermultiplet is:
\begin{equation}
S_{hyper}= -\int d\zeta^{(-4)}du \breve{q}^+ D^{++}q^+,
\end{equation}
where the integration is carried out over the analytic subspace.
The rules of harmonic integration are given in
\cite{gios}.

The ${\cal N}=2$ vector multiplet is described by  a real analytic
superfield $V^{++}(x, \theta^+, \bar{\theta}^+, u)$ with $U(1)$
charge equal to $+2$. It plays the role of the gauge connection in
the covariantized harmonic derivative $${\cal D}^{++}= D^{++}
+igV^{++},$$ so that the ``flat'' commutation relations with
$D^+_{(\alpha,\dot\alpha)}$ are preserved: $[{\cal D}^{++}, {\cal
D}^+_{\alpha,\dot\alpha}]=0.$ In the Wess-Zumino gauge the
component contents of $V^{++}$ is reduced to the off-shell ${\cal
N}=2$ vector multiplet. Namely this superfield $V^{++}$ is an
unconstrained prepotential for ${\cal N}=2$ SYM theory and all
other objects, for example the superfield strength ${\cal W}$, are
expressed in its terms \cite{gikos},
\cite{gios}.  The superfield strength is expressed
through non analytic superfield $V^{--}$ satisfying the equation
$D^{++}V^{--} -D^{--}V^{++}+i[V^{++}, V^{--}]=0$ . This equation
has the solution in form of the power series 
\begin{equation}
V^{--}=\sum^{\infty}_{n=1}\int d^{12}z du_1 ...du_n
(-i)^{n+1}\frac{V^{++}(z,u_1) ...
V^{++}(z,u_n)}{(u^+u^+_1)(u^+_1u^+_2) ... (u^+_n u^+)},
\end{equation}
which is a non-local functional of $V^{++}$ in harmonic sector.
The conditions of $u$ independence of ${\cal W}$, ${\cal
D}^{\pm\pm} {\cal W}=0$ lead to
\begin{equation}
{\cal W}=-\frac{1}{4}(\bar{\cal D}^+)^2 V^{--}, \bar{\cal
W}=-\frac{1}{4}({\cal D}^+)^2 V^{--}.
\end{equation}
The remaining properties of ${\cal W}, \bar{\cal W}$ are covariant
chirality (antichirality), ${\cal D}^{\pm}_\alpha \bar{\cal W} =
\bar{\cal D}^{\pm}_{\dot\alpha} {\cal W}=0$, and the Bianchi
identity: ${\cal D}^{\alpha i}{\cal D}^j_\alpha {\cal W}=\bar{\cal
D}^i_{\dot\alpha}\bar{\cal D}^{\dot\alpha j} \bar{\cal W}$. For
further use we will write down also the algebra of covariant
derivatives:
\begin{equation}\label{algebra}
\{{\cal D}^+_\alpha, {\cal D}^+_\beta\}=\{\bar{\cal
D}^+_{\dot\alpha}, \bar{\cal D}^+_{\dot\beta}\}=\{{\cal
D}^+_\alpha, \bar{\cal D}^+_{\dot\alpha}\}=0,
\end{equation}
$$ \{{\cal D}^+_\alpha, {\cal
D}^-_\beta\}=-2i\varepsilon_{\alpha\beta}\bar{\cal W}, \quad
\{\bar{\cal D}^+_{\dot\alpha}, \bar{\cal D}^-_{\dot\beta}\}=
2i\varepsilon_{\dot\alpha\dot\beta}{\cal W}, $$ $$ \{\bar{\cal
D}^+_{\dot\alpha}, {\cal D}^-_\alpha\}= -\{{\cal D}^+_\alpha,
\bar{\cal D}^-_{\dot\alpha}\} =2i{\cal D}_{\alpha\dot\alpha}, $$
$$ [{\cal D}^{++}, {\cal D}^-_\alpha]={\cal D}^+_\alpha, \quad
[{\cal D}^{++}, \bar{\cal D}^-_{\dot\alpha}]={\cal
D}^+_{\dot\alpha}, $$
\begin{equation}\label{A2}
[{\cal D}^{\pm}_{\alpha}, {\cal D}_{\beta \dot\beta}] =
\varepsilon_{\alpha \beta} \bar{D}^{\pm}_{\dot\beta} \bar{\cal
W},\quad [\bar{\cal D}^{\pm}_{\dot\alpha}, {\cal D}_{\beta
\dot\beta}] = \varepsilon_{\dot\alpha \dot\beta}
D^{\pm}_{\beta}{\cal W},
\end{equation}
$$[{\cal D}_{\alpha \dot\alpha}, {\cal D}_{\beta \dot\beta}] =
\frac{1}{2i}\{ \varepsilon_{\alpha \beta} \bar{D}^+_{\dot\alpha}
\bar{D}^-_{\dot\beta} \bar{\cal W} + \varepsilon_{\dot\alpha
\dot\beta} D^-_{\alpha}D^+_{\beta} {\cal W} \}.$$

Description of the ${\cal N}=4$ SYM theory in ${\cal N}=2$
harmonic superspace includes hypermultiplet and vector multiplet
superfields. The hypermultiplet superfield $q^+$ belongs to the
adjoint representation of the gauge group and is minimally coupled
to the vector multiplet. Action of ${\cal N}=4$ SYM theory in
the approach under consideration has the form:
\begin{equation}\label{01} S= \frac{1}{2g^2} \mbox{tr} \int d^8z
{\cal W}^2 -\frac{1}{2} \mbox{tr} \int d \zeta^{(-4)}
q^{+a}(D^{++} +iV^{++}) q^+_a.
\end{equation}
Here $a=1,2$ is the index of the Pauli-G\"{u}rsey rigid $SU(2)$
symmetry: $q^+_a = (q^+, \breve{q}^+), q^{+a}=\varepsilon^{ab}
q^+_a =(\breve{q}^+, -q^+),$ and $d^8z = d^4x d^2\theta^+
d^2\theta^- du$ is the chiral superspace integration measure.
Similarly, $d\zeta^{(-4)} = d^4x d^2\theta^+ d^2 \bar{\theta}^+ du$
is the analytic measure. The off-shell action (\ref{01}) allows us to
develop a manifest ${\cal N}=2$ supersymmetric quantization.
Moreover, this action is invariant under hidden on-shell extra
${\cal N}=2$ supersymmetry transformation \cite{gios} realized in
terms of the vector and hypermultiplet superfields as follows:
\begin{equation}\label{offsec}
\delta V^{++}=(\varepsilon^{\alpha a} \theta^+_a
+\bar{\varepsilon}^a_{\dot\alpha}\bar{\theta}^{+\dot\alpha})q^+_a,
\quad \delta q^+_a=-\frac{1}{2} (D^+)^4 [(\varepsilon^\alpha_a
\theta^-_\alpha +\bar{\varepsilon}_{\dot\alpha
a}\bar{\theta}^{-\dot\alpha})V^{--}].
\end{equation}
As a result, the model is (on-shell) ${\cal N}=4$ supersymmetric.

The on-shell structure of model is defined in terms of solutions
to the corresponding equation of motion
\begin{equation}
D^{++}q^{+a} +ig[V^{++}, q^{+a}]=0, \quad ({\cal D}^{+})^2{\cal W}
=[q^{+a}, q^+_a].
\end{equation}
The simplest solution to these equations in Abelian case  forms a set of constant
background fields, which transform linearly through each
other mixing up the ${\cal W}, \bar{\cal W}$ with $q^+_a$ under hidden
${\cal N}=2$ supersymmetry transformation \cite{bbku},
\cite{bbiko}:
\begin{equation}\label{001}
\delta{\cal W}=\frac{1}{2} \bar{\varepsilon}^{\dot\alpha
a}\bar{D}^-_{\dot\alpha} q^+_a, \quad \delta\bar{\cal
W}=\frac{1}{2} {\varepsilon}^{\alpha a}{D}^-_{\alpha} q^+_a,
\end{equation}
$$
\delta q^+_a= \frac{1}{4} (\varepsilon^{\alpha}_a
D^+_{\alpha}{\cal W} +\bar{\varepsilon}^{\dot\alpha}_a
\bar{D}^+_{\dot\alpha}\bar{\cal W}).
$$

Generic vacuum of any ${\cal N}=2$ superconformal models, like for
example ${\cal N}=4$ SYM theory,  includes only massless $U(1)$
vector multiplets and massless neutral hypermultiplets, since
charged hypermultiplets get masses by means of the Higgs
mechanism. The manifold of vacua is determined by the conditions
of vanishing scalar potential (F-flatness plus D-flatness). The
set of vacua with only massless neutral hypermultiplets forms the
"Higgs branch" of the theory. The set with only $U(1)$ vector
multiplets forms the  "Coulomb branch". At last, the set of vacua
with both kinds of the multiplets forms the "mixed branch". Thus,
the low energy fields propagating on the mixed branch are massless
neutral scalars, spinors and $U(1)$ vectors which form the on
shell superfields ${\cal W}, \bar{\cal W}, q^{+a}, q^{-a}$
possessing the properties $$ (D^{\pm})^2{\cal
W}=(\bar{D}^{\pm})^2\bar{\cal W}=0, $$ $$
D^{++}q^{+a}=D^{--}q^{-a}=0,\quad q^{-a}\equiv D^{--}q^{+a},\quad
D^-_\alpha q^{-a}=\bar{D}^-_{\dot\alpha}q^{-a}=0.$$ Further we
will consider the low-energy effective action in ${\cal N}=4$ SYM
theory just on the mixed branch in the above sense. On the other
hand, since the ${\cal N}=2$ vector multiplet and the
hypermultiplet form the ${\cal N}=4$ vector multiplet in the case
under consideration, we can treat the vacuum state as an unique
Coulomb branch of the model. We want to emphasize that use of the
various terms for vacuum state in this theory is a matter of
convention.

The manifestly ${\cal N}=2$ supersymmetric Feynman rules in
harmonic superspace have been developed in \cite{gikos} (see also
\cite{gios}, \cite{oh}). Calculations of quantum
corrections may contain  potentially dangerous harmonic
singularities, that is harmonic distributions at coinciding
points. The problem of coinciding harmonic singularities in the
framework of harmonic supergraph Feynman rules was first discussed
in \cite{singul} where some solution to the problem was
considered. The background field method for constructing the
effective action in harmonic superspace has been developed in
Refs. \cite{backgr}, \cite{bbiko} (see also \cite{town} for
construction of the background field method in standard $N=2$
superspace). This method allows us to find the effective action for
arbitrary ${\cal N}=2$ supersymmetric gauge model in a form
preserving the manifest ${\cal N}=2$ supersymmetry and classical
gauge invariance in quantum theory. In framework of background
field method the fields $V^{++}, q^{+}$ are split into
classical $V^{++}, q^{+}$ and quantum $v^{++}, Q^{+}$ fields with
imposing the gauge conditions only on quantum fields. The Feynman
rules are based on the quantum action $S_{quant}$ of the form
$S_{quant} = S_{2} + S_{int}$ where  $S_{2}$ is a quadratic
form in quantum fields and ghosts and $S_{int}$ describes
interaction. Both action $S_{2}$ and action $ S_{int}$ depend on
background fields. All details are given in \cite{backgr},
\cite{bbiko}.

The action $S_{2}$ defines the propagators depending on background
fields. Further we use the background covariant gauge ${\cal
D}^{++}v^{++}=0$. In this case the propagator of quantum gauge
superfield has the form
\begin{equation}\label{fgvlong}
G^{(2,2)}(1,2)=\frac{1}{2{\stackrel{\frown}{\Box}}_1
{\stackrel{\frown}{\Box}}_2} ({\cal D}^+_1)^4 ({\cal D}^+_2)^4
\{\delta^{12}(z_1-z_2) (D^{--}_2)^2\delta^{(-2,2)}(u_1,u_2)\}.
\end{equation}
We emphasize that this propagator is analytic superfield in each
argument. The propagator of the Faddeev-Popov ghosts $b$ and
$c$ is written as follows
\begin{equation}\label{fgom}
G^{(0,0)} (1,2) =i<b (1) c^T (2)>=
\frac{1}{{\stackrel{\frown}{\Box}}_1}({\cal D}^+_1)^4 ({\cal
D}^+_2)^4
\{\delta^{12}(z_1-z_2)\frac{(u^-_1u^-_2)}{(u^+_1u^+_2)^3} \}.
\end{equation}
The $q^+$ hypermultiplet propagator associated with
the action (\ref{01}) for external $V^{++}$ has the form
\begin{equation}\label{fgq}
G^{(1,1)}(1,2)=i<q^{+a}(1) \breve{q}^+_b (2)> =-\delta^a_b
\frac{1}{{\stackrel{\frown}{\Box}}_1}({\cal D}^+_1)^4 ({\cal
D}^+_2)^4 \{\delta^{12}(z_1-z_2)\frac{1}{(u^+_1u^+_2)^3} \}.
\end{equation}

The propagators contain the operator ${\stackrel{\frown}{\Box}}=-\frac{1}{2} ({\cal
D}^{+})^4({\cal D}^{--})^2$ which transforms each covariantly analytic
superfield into a covariantly analytic one.  On the space
of such superfields, the operator ${\stackrel{\frown}{\Box}}$ is equivalent to
the second-order Laplace like differential operator
\begin{equation}\label{smilebox}
{\stackrel{\frown}{\Box}} = \frac{1}{2} {\cal
D}^{\dot\alpha\alpha}{\cal D}_{\alpha\dot\alpha} + \frac{i}{2}
({\cal D}^{+\alpha}{\cal W}){\cal D}^-_\alpha + \frac{i}{2}
(\bar{\cal D}^+_{\dot\alpha} \bar{\cal W}) \bar{\cal
D}^{-\dot\alpha} - \frac{i}{4} (\bar{\cal D}^+_{\dot\alpha}
\bar{\cal D}^{+\dot\alpha} \bar{\cal W}) {\cal D}^{--}
\end{equation}
$$
+ \frac{i}{8}[{\cal D}^{+\alpha},{\cal D}^-_\alpha] {\cal W} +
 \frac{1}{2}\{{\cal W},\bar{\cal W}\}.
$$
This is  a consequence of algebra of covariant derivatives
(\ref{algebra}). It is remarkable that the differential part of
${\stackrel{\frown}{\Box}}$ is uniquely determined from the
requirements that (i) ${\stackrel{\frown}{\Box}}$ a constructed in
terms of the covariant derivatives only; (ii)
${\stackrel{\frown}{\Box}}$ transforms  every covariantly analytic
superfield into a covariantly analytic one. This  operator
is said to be the analytic d'Alambertian. Among the important
properties of ${\stackrel{\frown}{\Box}}$ is \cite{backgr} :
$[{\cal D}^+_{(\alpha,\dot\alpha)}, {\stackrel{\frown}{\Box}}] =
0.$ The ${\cal N}=2$ propagators have a complicated structure due
to nontrivial dependence on harmonics. Fortunately, as is shown
recently \cite{exprop}, the harmonic dependence of the ${\cal N}=2$
propagators simplifies drastically if the background vector
multiplet satisfies the classical equations of motion ${\cal
D}^{ij}{\cal W}=\bar{\cal D}^{ij}\bar{\cal W}=0.$ In this case the
harmonic dependence of the propagators is completely factorization
what helps to keep the harmonic dependence of ${\cal N}=2$
supergraphs under control.

Evaluation of effective action within the background field method is
accompanied often by use of proper time or heat kernel techniques.
These techniques allow one to sum up efficiently an infinite set of
Feynman diagrams with increasing number of insertions of the
background fields and to develop a background field derivative
expansion of the effective action in manifestly gauge covariant
way. The background field method and heat kernel techniques for
${\cal N}=1$ SYM theories were well-developed (see \cite{24},
\cite{25} for reviews). The background field method in harmonic
superspace was elaborated in Refs. \cite{backgr}, some of its
important applications were reviewed in \cite{bbiko}. However,
until recently, the many aspects of heat kernel techniques in
${\cal N}=2$ superspace remained almost totally unstudied\footnote{Some
new results were presented in \cite{2}.}. In Section 5,
we use the ${\cal N}=2$ heat kernel techniques to construct
the one-loop effective action of ${\cal N}=4$ SYM theory in ${\cal
N}=2$ harmonic superspace and extend the one-loop results of
\cite{1}, \cite{1b} for the leading low-energy quantum corrections
to the next-to-leading contributions to the effective action depending
both on ${\cal N}=2$ vector multiplet and on ${\cal N}=2$
hypermultiplet.


\section{One-loop effective action in hypermultiplet sector}


We consider the ${\cal N}=4$ SYM theory with gauge group $SU(2)$
formulated in ${\cal N}=2$ harmonic superspace. Action of the
model has the form (\ref{01}). On the mixed branch the gauge group
$SU(2)$ is broken down to $U(1)$ so that background fields
$V^{++}, q^{+}$ lie in the Cartan subalgebra of the gauge group.
We start, like in \cite{1b}, with carrying out background-quantum
splitting by the rule $q^{+a}\rightarrow q^{+a} +Q^{+a}, \quad
V^{++}\rightarrow V^{++} +gv^{++}$. Here $q^{+a}, V^{++}$ are the background
fields and $Q^{+a}, v^{++}$ are the quantum ones. For one-loop calculations it
is sufficient to consider only the part of quantum action $S_{quant}$ which is quadratic
in quantum superfields:
\begin{equation}\label{02}
S_2 = -\frac{1}{2} \mbox{tr} \int d \zeta^{(-4)} [v^{++}
{\stackrel{\frown}{\Box}} v^{++} + Q^{+a}(D^{++} +i V^{++})Q^+_a +
Q^{+a}(ig v^{++})q^+_a +q^{+a} (ig v^{++})Q^+_a] + \dots.
\end{equation}
Where $...$ means the ghost contribution.  The operator
${\stackrel{\frown}{\Box}}$ includes the background ${\cal N}=2$
superfield strengths ${\cal W}, \bar{\cal W}$ which  acts in the
adjoint representation of the gauge group and has the form
(\ref{smilebox}). The set of the background superfields is
on-shell, $({\cal D}^+)^2{\cal W} =0, \quad {\cal D}^{++}q^{+a}=0$
and satisfies the relations ${\cal D}^{+\alpha}{\cal D}^-_{\alpha}
{\cal W}=0$ and  ${\cal D}^{\pm\alpha}{\cal D}^{\pm}_{\alpha}
{\cal W}=0.$  We redefine $g \,Q^+ \rightarrow Q^+$ and write the
background superfields $V^{++}=\tau_3V^{++}_3$,
$q^{+}=\tau_3q^{+}_3$ and hence ${\cal W}=\tau_3{\cal W}_3.$ Here
$\tau_i =\frac{1}{\sqrt{2}}\sigma_i$ are generators of $su(2)$
algebra: $$[\tau_i, \tau_j] =i\sqrt{2}\epsilon_{ijk} \tau_k ,
\quad \mbox{tr}(\tau_i \tau_j) = \delta_{ij}. $$ For the
background belonging to an Abelian subgroup we have also the
further restrictions: ${\cal D}^{\pm \alpha}{\cal W} = D^{\pm
\alpha}{\cal W}$, and similarly for $\bar{\cal W}$ with $D,
\bar{D}$ being ``flat'' derivatives. Then taking into account the
on-shell conditions and that all quantum superfields are in the
adjoint representation $v^{++} =v^{++}_i \tau_i $, $Q^{+a}
=Q^{+a}_i \tau_i$ one gets
\begin{equation}\label{04}
{\stackrel{\frown}{\Box}}v^{++} = \Box_{cov}v^{++} +
\frac{i}{2}[D^{+\alpha} {\cal W}, {\cal D}^-_{\alpha}v^{++}]
+\frac{i}{2}[\bar{D}^{+ \dot\alpha}\bar{\cal W}, \bar{\cal
D}^-_{\dot\alpha} v^{++}] + [{\cal W },[\bar{\cal W}, v^{++}]].
\end{equation}
Here $\Box_{cov} = \frac{1}{2}{\cal D}^{\alpha\dot\alpha}{\cal
D}_{\alpha\dot\alpha},$ $q^{+a} v^{++} Q^+_a = q^{+a}[v^{++},
Q^+_a].$ We also have $ {\cal D}^{++}q^1= D^{++}q^1 +
\sqrt{2}V^{++} q^2,$ ${\cal D}^{++}q^2= D^{++}q^2 - \sqrt{2}V^{++}
q^1,$ ${\cal D}^{++}q^3 = D^{++} q^3,$ ${\cal D}_m =D_m
+\sqrt{2}A_m. $

Thus, the quadratic part of the action is written  as
\begin{equation}\label{05}
S_2 = -\frac{1}{2}\int d \zeta^{(-4)} \{v^{++}_1(\Box_{cov}
+2{\cal W}\bar{\cal W})v^{++}_1 + v^{++}_2(\Box_{cov} +2{\cal
W}\bar{\cal W})v^{++}_2
\end{equation}
$$+\frac{1}{\sqrt{2}}v^{++}_1((D^{+\alpha}{\cal W})D^-_{\alpha} +
(\bar{D}^+_{\dot\alpha}\bar{\cal W}) \bar{D}^{-
\dot\alpha})v^{++}_2 -\frac{1}{\sqrt{2}}v^{++}_2((D^{+\alpha}{\cal
W})D^-_{\alpha} + (\bar{D}^+_{\dot\alpha}\bar{\cal W}) \bar{D}^{-
\dot\alpha})v^{++}_1$$ $$+ v^{++}_3 \Box v^{++}_3 +Q^{+a}_i
D^{++}Q^+_{ia} +Q^{+a}_1(\sqrt{2}V^{++})Q^+_{2a} +
Q^{+a}_2(-\sqrt{2}V^{++})Q^+_{1a}$$ $$+
v^{++}_2\sqrt{2}(q^{+a}Q^+_{a1}-Q^{+a}_1q^+_a) + v^{++}_1\sqrt{2}
(Q^{+a}_2 q^+_a -q^{+a} Q^{+}_{a2})\} + ...$$
Here and further we
suppress the index 3 at background superfields, .... means the
ghost contribution. It can be seen from (\ref{05}) that only the
components of quantum superfields carrying indices $1, 2$ have a
non-trivial background-dependent propagator, while the quantum
superfield components with index 3 do not interact with the
background and totally decouple.

We define the new complex quantum superfields $$
\chi^{++}=\frac{1}{\sqrt{2}}(v^{++}_1 + iv^{++}_2), \quad
\bar{\chi}^{++}=\frac{1}{\sqrt{2}}(v^{++}_1 - iv^{++}_2), $$
$$\eta^{+a}=\frac{1}{\sqrt{2}}(Q^{+a}_1 +iQ^{+a}_2), \quad
\bar{\eta}^{+a}=\frac{1}{\sqrt{2}}(Q^{+a}_1 - iQ^{+a}_2).$$ As a
result, the action (\ref{05}) takes  the form
\begin{equation}\label{06}
S_2=-\int d \zeta^{(-4)} \{\bar{\chi}^{++}(\Box_{cov} +2{\cal
W}\bar{\cal W} -\frac{i}{\sqrt{2}}((D^{+\alpha} {\cal
W})D^-_{\alpha} + (\bar{D}^+_{\dot\alpha}\bar{\cal
W})\bar{D}^{-\dot\alpha}))\chi^{++}
\end{equation}
$$+\bar{\eta}^{+a}(D^{++} - i\sqrt{2} V^{++})\eta^+_a -i\chi^{++}
\sqrt{2} q^{+a} \bar{\eta}^+_a + i\bar{\chi}^{++} \sqrt{2}q^{+a}
\eta^+_a + \frac{1}{2}v^{++}_3 \Box v^{++}_3  +\frac{1}{2}
Q^{+a}_3D^{++}Q^+_{3a}\}.$$ This form of action is very convenient
for perturbative calculations. It has a diagonal part $S_{0}$
that defines the propagators and a non-diagonal part $V$ responsible for
interaction. The non-interacting fields $Q^{a+}_3, v^{++}_3$ will
be omitted. As the next step, we introduce the operator
\begin{equation}\label{07}
{\stackrel{\frown}{\Box}}_{short} =\Box_{cov} +2{\cal W}\bar{\cal
W} -\frac{i}{\sqrt{2}}((D^{+\alpha}{\cal W}){\cal D}^-_{\alpha} +
(\bar{D}^+_{\dot\alpha}\bar{\cal W})\bar{\cal D}^{- \dot\alpha}),
\end{equation}
which is obtained from (\ref{smilebox}) by keeping only the
superfields ${\cal W}, \bar{\cal W}$ associated with the unbroken
$U(1)$ subgroup and putting them on shell. Note that on shell the
form of the operator (\ref{smilebox}) and (\ref{07}) is related to
each other by ${\cal W} \rightarrow \frac{{\cal W}}{-\sqrt{2}},$
$\bar{\cal W} \rightarrow \frac{\bar{\cal W}}{-\sqrt{2}}.$ Throughout
the paper, we will still be using on shell notation (\ref{smilebox}) for
(\ref{07}). Now it is easy to read off the Feynman rules for
calculating the effective action, which are derived in a similar
manner to \cite{gikos}, \cite{gios}, \cite{backgr}. For the propagators
of the gauge fields
$\chi^{++}$ and $\bar{\chi}^{++}$ and the hypermultiplets $\eta^+_a$
and $\bar{\eta}^{+a}$,  we use (\ref{fgvlong}) and (\ref{fgq})
respectively. Vertices are taken directly from the $V$ in the form
$$ V  = -i\chi^{++} \sqrt{2} q^{+a} \bar{\eta}^+_a +
i\bar{\chi}^{++} \sqrt{2}q^{+a} \eta^+_a .$$ The Feynman rules
look standard. We emphasize only the important point. In each
vertex which includes the integral over analytic subspace one can
use the factor $(D^+)^4$ from one propagator and convert the
integral over $d \zeta^{(-4)}$ to integral over full ${\cal N}=2$
measure $d^{12} z.$

We point out that the functional change of variables in
(\ref{06}): $$ \chi^{++}(1)\rightarrow \chi^{++}(1) -i\int
d\zeta^{(-4)}_2 G^{(2,2)}(1|2) q^{+a}(2)\eta^+_a(2), $$ $$
\bar{\chi}^{++}(1)\rightarrow \bar{\chi}^{++}(1) +i\int
d\zeta^{(-4)}_2 G^{(2,2)}(1|2) q^{+a}(2)\bar{\eta}^+_a(2), $$ with
$G^{2,2}(1|2)$ (\ref{fgvlong}) leads to diagonalization of the
operator $$ \int d\zeta^{(-4)}_1d\zeta^{(-4)}_2
\bar{\eta}^{+a}(1)\{\delta^b_a {\cal D}^{++}_1\delta^{(1,3)}(1|2)
+q^+_a(1)G^{(2,2)}(1|2)q^{+b}(2)\}\eta^+_b(2). $$ Then, the
one-loop effective action $\Gamma[V^{++}, q^+]$ defined by the
path integral $$ e^{i\Gamma[V^{++}, q^+]}=\int {\cal
D}\bar{\eta}^{+}{\cal D}\eta^+ {\cal D}\bar{\chi}^{++}{\cal
D}{\chi}^{++}e^{i S_2 [\eta^+,
\bar{\eta}^{+},\bar{\chi}^{++},{\chi}^{++}, V^{++}, q^+ ]} $$ can
be formally written as
\begin{equation}\label{log}
\Gamma[V^{++}, q^+, \breve{q}^+]=i \mbox{Tr} \ln \{\delta^b_a
{\cal D}^{++}_1\delta^{(1,3)}(1|2)
+q^+_a(1)G^{(2,2)}(1|2)q^{+b}(2)\} +\Gamma[V^{++}],
\end{equation}
where the last term $\Gamma[V^{++}]$ is a part of the full one-loop
effective action which depends only on ${\cal N}=2$ gauge
superfield. We will study mostly the first term in the above equation since it
includes all the hypermultiplet dependence. The expression
(\ref{log}) written as an analytic nonlocal superfunctional will
be a starting point for our calculations of the one-loop effective action
in the hypermultiplet sector. Eq. (\ref{log}) shows that the
effective action is well defined in  perturbation theory in
powers of the non-local interaction
$q^{+}_a(1)G^{(2,2)}(1|2)q^{+b}(2).$ It leads to effective action
in the form $\Gamma[V^{++}, \breve{q}^+, q^+]= \sum_{n=1}^{\infty}
\Gamma_{2n} [V^{++}, \breve{q}^+, q^+].$  Here the $2n$-th term is
given by a supergraph with $2n$ external $\breve{q}^+, q^+$-legs
and any number $V^{++}$-legs. Since $\Gamma[V^{++}, \breve{q}^+,
q^+]$ is gauge invariant by construction, one can expect that each
coefficient $\Gamma_{2n}$ depends on background superfield
$V^{++}$ only via the strengths ${\cal W}, \bar{\cal W}$ and their
covariant derivatives. We emphasize that the supergraphs
associated with this procedure contain the background dependent
superpropagators.


\section{Analysis of supergraphs for
hypermultiplet dependent contributions to effective action}


The hypermultiplet dependent contributions to the one-loop
effective action are presented by following infinite sequence of
the supergraphs: \vspace*{5mm}

\hspace*{.5cm} \Lengthunit=0.95cm \GRAPH(hsize=3){\mov(0,-.5){
\halfcirc(1.0)[u]\halfwavecirc(1.0)[d]
\mov(-.6,0){\lin(-.5,0)}\mov(.45,0){\lin(.5,0)}}
\ind(18,-5){+}\ind(12,-3){2}\ind(-11,-3){1} }
\GRAPH(hsize=3){\lin(-.7,.7)\wavelin(1.1,0)\lin(0,-1)
\mov(1,-1){\lin(.7,-.7)\lin(0,1)\wavelin(-1.1,0)}\ind(19,-5){+}
\mov(.9,0){\lin(.7,.7)}\mov(-.2,-1){\lin(-.7,-.7)}
\ind(16,4){3}\ind(16,-15){4}\ind(-12,4){2}\ind(-12,-15){1} }
\hspace*{.5cm}
\GRAPH(hsize=3){\lin(-.7,.7)\wavelin(.6,0)\lin(0,-1)
\mov(.5,0){\lin(0,1)\lin(.45,0)}
\mov(1,-1){\lin(.7,-.7)\wavelin(0,1)\lin(-.5,0)\ind(13,4){+}
\mov(-.6,0){\lin(0,-1)\wavelin(-.6,0)}}
\mov(.9,0){\lin(.7,.7)}\mov(-.2,-1){\lin(-.7,-.7)}
\ind(16,4){4}\ind(16,-15){5}\ind(-12,4){2}\ind(-12,-15){1}
\ind(4,8){3}\ind(4,-18){6} } \hspace*{0.5cm}
\GRAPH(hsize=3){\mov(.1,0){\lin(-.7,.7)\wavelin(.55,0)}
\lin(0,-.5)\mov(0,-.5){\wavelin(0,-.5)\lin(-.7,0)}
\mov(.5,0){\lin(0,1)\lin(.5,0)}
\mov(1,-1){\lin(.7,-.7)\lin(0,.5)\wavelin(-.5,0)
\mov(-.6,0){\lin(0,-1)\lin(-.6,0)}}
\mov(.9,-.5){\wavelin(0,.5)\lin(.7,0)}
\mov(.8,0){\lin(.7,.7)}\mov(-.3,-1){\lin(-.7,-.7)}
\ind(16,5){4}\ind(16,-15){6}\ind(-13, 3){2}\ind(-12,-15){8}
\ind(3,8){3}\ind(4,-18){7}\ind(16,-5){5}\ind(-15,-5){1}
\ind(21,-5){+}\ind(27,-5) {\ldots} } \vspace*{4mm}

\noindent Here the wavy line stands for ${\cal N}=2$ gauge
superfield propagator and solid external and internal lines stand
for background hypermultiplet superfields and quantum
hypermultiplet propagators respectively. The numbers $1,2, ...$
mark the arguments $\zeta_A, u$ of the external hypermultiplets lines.
As we emphasized above the whole dependence of the
contributions on the background gauge superfield
is included into background-dependent propagators.

An arbitrary supergraph with $2n$ external hypermultiplet lines
looks like a ring consisting of $n$ links of the form $<\bar{\eta}^+
\eta^+ > <\chi^{++}\bar{\chi}^{++}>$ or $n$ links of the form $<
\eta^+ \bar{\eta}^+> <\bar{\chi}^{++}\chi^{++}>.$ The total
contribution of these two kinds of the $2n$-point supergraph is
given by following general expression
\begin{equation}\label{2n}
i \Gamma_{2n}=\frac{4}{n}\int
d\zeta_1^{(-4)}d\zeta_2^{(-4)}...d\zeta_{2n}^{(-4)}
\end{equation}
$$ \times \quad \frac{({\cal D}_1^+)^4({\cal D}_2^+)^4}
{(u^+_1u^+_2)^3}
\{\frac{1}{{\stackrel{\frown}{\Box}}_1}\delta^{12}_{1|2}\}
\frac{({\cal D}^+_2)^4({\cal
D}^+_3)^4}{{\stackrel{\frown}{\Box}}_2{\stackrel{\frown}{\Box}}_3}
({\cal D}^{--}_3)^2
\{\delta^{12}_{2|3}\delta^{(-2,2)}(u_2,u_3)\}$$ $$ \times \quad
\frac{({\cal D}_3^+)^4({\cal D}_4^+)^4}{(u^+_3u^+_4)^3}
\{\frac{1}{{\stackrel{\frown}{\Box}}_3}\delta^{12}_{3|4}\} \ \
.\,.\,. \ \ \frac{({\cal D}^+_{2n-2})^4({\cal
D}^+_{2n-1})^4}{{\stackrel{\frown}{\Box}}_{2n-2}{\stackrel{\frown}{\Box}}_{2n-1}}({\cal
D}^{--}_{2n-1})^2\{\delta^{12}_{2n-2|2n-1}
\delta^{(-2,2)}(u_{2n-2},u_{2n-1})\} $$ $$ \times \quad
\frac{({\cal D}^+_{2n-1})^4({\cal
D}^+_{2n})}{(u^+_{2n-1}u^+_{2n})^3}\{\frac{1}{{\stackrel{\frown}{\Box}}_{2n-1}}\delta^{12}_{2n-1|2n}\}
\frac{({\cal D}^+_{2n})^4({\cal
D}^+_1)^4}{{\stackrel{\frown}{\Box}}_{2n}{\stackrel{\frown}{\Box}}_1}({\cal
D}^{--}_1)^2\{\delta^{12}_{2n|1}\delta^{(-2,2)}(u_{2n},u_1)\} $$
$$ \times \quad q^+_a(z_1,u_1) q^{+a}(z_2,u_2)q^+_b(z_3,u_3)\ \
.\,.\,. \ \ q^+_c(z_{2n-1},u_{2n-1})q^{+c}(z_{2n},u_{2n}). $$ Here
and further, to avoid coincident harmonic singularities, we keep
the ${\cal N}=2$ gauge superfield propagator in the form which is
manifestly analytic in both argument \cite{singul}, \cite{gios}.

The factor $4/n$ has the following origin (see \cite{1b}). The
contribution from the ring type supergraph composed from $n$
repeating links $<\bar{\eta}^+ \eta^+ >
<\chi^{++}\bar{\chi}^{++}>$ appears with the symmetry factor
$2/n.$ The same factor $2/n$ arises from the supergraph composed
from $n$ repeating links $< \eta^+ \bar{\eta}^+>
<\bar{\chi}^{++}\chi^{++}>.$ Further, each vertex brings the
factor $-i$, every $< \eta^+ \bar{\eta}^+>$ and
$<\bar{\chi}^{++}\chi^{++}>$ propagators contribute the factor $i$
and $i/2$ respectively. Hence total of $n$ links contributes
$2^{-n}.$ Any vertex also carries the coefficient $\sqrt{2}.$ This
leads to the total factor $2^n.$ Substituting all these
contributions together, we obtain just the coefficient $4/n.$

We begin with direct calculation of the term
$\Gamma_2[V^{++}, q^{+a}]$ which in the analytic basis reads
\begin{equation}\label{0}
i\Gamma_2 =\int d \zeta^{(-4)}_1 d\zeta^{(-4)}_2 du_1 du_2
\{\frac{({\cal D}^+_1)^4({\cal D}^+_2)^4}{(u^+_1 u^+_2)^3}
\frac{1}{{\stackrel{\frown}{\Box}}_1}\delta^{12}(1|2)\}
\end{equation}
$$ \times \{\frac{({\cal D}^+_2)^4({\cal
D}^+_1)^4}{{\stackrel{\frown}{\Box}}_2{\stackrel{\frown}{\Box}}_1}\delta^{12}(2|1)
({\cal D}^{--}_1)^2 \delta^{-2,2}(u_2,u_1)\} \quad q^+_a(z_1,u_1)
q^{+a}(z_2,u_2). $$ According to the general strategy of handling
such supergraphs we should first to restore the full Grassmann
integration measure at the vertices by rule $d^{12}z_1\, d^{12}z_2
= d^{(-4)}\zeta_1 \, d^{(-4)}\zeta_2  (D^+_1)^4 \,(D^+_2)^4.$
Since we are interested in contributions to the effective action which do not depend on
space-time derivatives ${\cal D}_m q^+_a$ and spinor derivatives ${\cal
D}^{-}_{(\alpha,\dot\alpha)}q^+_a$ of the background hypermultiplets,
it is sufficient to impose the following superfields
restrictions
\begin{equation}\label{rest}
{\cal D}_{(\alpha,\dot\alpha)}^-q^+_a =0, \quad {\cal
D}_{(\alpha,\dot\alpha)}^+ q^{-a}=0.\end{equation} Integrating by
parts with the help of delta function we shrink a loop into a
point in superspace. However, in harmonic superspace there still
remains a non-local expression
\begin{equation}\label{G2}
i\Gamma_2=\int \frac{dz du_1 du_2}{(u^+_1 u^+_2)^3} \frac{({\cal
D}^+_2)^4({\cal D}^+_1)^4}{{\stackrel{\frown}{\Box}}^2_1
{\stackrel{\frown}{\Box}}_2} \delta^{12}(z)|\{({\cal D}^{--}_1)^2
\delta^{-2,2}(u_2,u_1)\} q^+_a(z_1,u_1)q^{+a}(z_1,u_2).
\end{equation}
We point out that the presence of the harmonic distribution $(u^+_1
u^+_2)^{-3}$ with coincident singularities. If we had the
flat covariant derivatives we could use the important identity:
$(D^+_1)^4(D^+_2)^4 \delta^8(\theta_1-\theta_2)=(u^+_1u^+_2)^4$
and would get a simplification. However, the expression under consideration
(\ref{G2}) contains the covariant spinor derivatives and further
analysis becomes more complicated. In principle, one can use
the idea of ref. \cite{2} to express the covariant derivatives
${\cal D}^{+(\alpha, \dot\alpha)}_2$ through the covariant
derivatives ${\cal D}^{+(\alpha, \dot\alpha)}_1$ to evaluate the
two-point function of the form $ ({\cal D}^+_1)^4 ({\cal D}^+_2)^4
\delta^{12}(z_1 -z_2)\frac{1}{(u^+_1 u^+_2)^q}.$ It can help to
calculate the $\Gamma_2$ (\ref{G2}), however a procedure of such a
calculation still looks very complicated technically. From our
point of view, it is more convenient to start with the
representation (\ref{0}) and work in the analytic subspace
\cite{2loop}. Then instead of computing the contribution
(\ref{G2}) in full ${\cal N}=2$ superspace we should actually look
for an equivalent expression of  the form $$ \int
d\zeta^{-4}(D^+)^4 {\cal L}({\cal W}, \bar{\cal W}, q^+) $$ with
some ${\cal L}.$ To do that we return in (\ref{G2}) to analytic
subspace, use twice the relation \cite{2} and apply the harmonic
identities $(u^+_1 u^+_2 )|_{1=2}=0$, $D^{--}_1(u^+_1
u^+_2)=(u^-_1 u^+_2)$, $(u^+_1 u^-_2)|_{1=2}=1$
\begin{equation}\label{D}
\frac{1}{(u^+_1 u^+_2)^3}({\cal D}^+_1)^4 ({\cal D}^+_2)^4({\cal
D}^+_1)^4 = ({\cal D}^+_1)^4\{...+ (u^+_1 u^+_2)(u^-_2
u^+_1)^2(u^-_1 u^+_2)^2 {\stackrel{\frown}{\Box}}_1
{\stackrel{\frown}{\Box}}_2 + ...\}.
\end{equation}
As a result ones get
\begin{equation}
i\Gamma_2 =\int d\zeta^{(-4)} du_1du_2 (u^+_1u^+_2)
(q^+_a(u_1)q^{+a}(u_2))(D^{--}_1)^2 \delta^{(-2,2)}(u_2u_1)
\frac{({\cal D}^+)^4}{{\stackrel{\frown}{\Box}}_2}\delta^{12}(z)|.
\end{equation}
In order to be able to take the factor $(D^{--})^2$ off the
harmonic delta function and apply the harmonic identities
$$
(D^{--}_1)^2 \delta^{(-2,2)}(u_2u_1)=(D^{--}_2)^2
\delta^{(2,-2)}(u_2u_1),\quad D^{--}q^+=q^-,\quad
(u^+_1u^+_2)|_{1=2}=0,  $$ we temporarily restore the full measure
of integration. After that we return to the analytic measure. In
doing this we keep in mind the restrictions (\ref{rest}).
Integration over one set of harmonics leads to final result for
${\Gamma}_2$
\begin{equation}
i\Gamma_2=\int d\zeta^{(-4)}du (-4 q^-_aq^{+a})\frac{({\cal
D}^+)^4}{{\stackrel{\frown}{\Box}}}\delta^{12}(z)|.
\end{equation}

The next step is to calculate  the four-leg contribution
$\Gamma_4 [q^+].$ We start with general relation (\ref{2n}) for
$n=2$ and perform the same manipulations as in preliminary case.
It gives
\begin{equation}\label{a1}
i\Gamma_{4}=\int d\zeta^{(-4)}_1...d\zeta^{(-4)}_4 du_1 ...du_4
\frac{({\cal D}_1^+)^4({\cal D}_2^+)^4} {(u^+_1u^+_2)^3}
\{\frac{1}{{\stackrel{\frown}{\Box}}_1}\delta^{12}(1|2)\}
\end{equation}
$$\times [\frac{({\cal D}_2^+)^4({\cal
D}_3^+)^4}{{\stackrel{\frown}{\Box}}_2
{\stackrel{\frown}{\Box}}_3} \delta^{12}(2|3)({\cal
D}^{--}_3)^2\delta^{(-2,2)}(u_2,u_3)] \frac{({\cal D}_3^+)^4({\cal
D}_4^+)^4} {(u^+_3u^+_4)^3}
\{\frac{1}{{\stackrel{\frown}{\Box}}_3}\delta^{12}(3|4)\} $$
$$\times [\frac{({\cal D}_4^+)^4({\cal
D}_1^+)^4}{{\stackrel{\frown}{\Box}}_4
{\stackrel{\frown}{\Box}}_1} \delta^{12}(4|1)({\cal
D}^{--}_1)^2\delta^{(-2,2)}(u_4,u_1)] q^+_a(z_1,u_1)
q^{+a}(z_2,u_2) q^+_b(z_3,u_3) q^{+b}(z_4,u_4)= $$ $$ =\int
\frac{dz du_1 ... du_4}{(u^+_1 u^+_2)^3(u^+_3 u^+_4)^3}
\frac{({\cal D}^+_1)^4 ({\cal D}^+_2)^4 ({\cal D}^+_3)^4 ({\cal
D}^+_4)^4}{{\stackrel{\frown}{\Box}}_1^2
{\stackrel{\frown}{\Box}}_2 {\stackrel{\frown}{\Box}}_3^2
{\stackrel{\frown}{\Box}}_4}\delta^{12}(z)| ({\cal D}^{--}_3)^2
\delta^{(-2,2)}(u_2,u_3)$$$$\times ({\cal D}^{--}_1)^2
\delta^{(-2,2)}(u_4,u_1) q^+_a(1) q^{+a}(2) q^+_b(3)q^{+b}(4). $$
Using identities $({\cal D}^{--}_3)^2
\delta^{(-2,2)}(u_2,u_3)=({\cal D}^{--}_2)^2
\delta^{(2,-2)}(u_2,u_3)$,\\ and $({\cal D}^+_2)^4({\cal
D}^{--}_2)^2 ({\cal D}^+_3)^4 \delta^{(2,-2)}(u_2,u_3) =-2
{\stackrel{\frown}{\Box}}_2 ({\cal D}^+_2)^4 $ one obtains $$
i\Gamma_4 = \int dz \frac{du_1 du_2}{(u^+_1 u^+_2)^6} \frac{({\cal
D}^+_1)^4 ({\cal
D}^+_2)^4}{{\stackrel{\frown}{\Box}}_1^2{\stackrel{\frown}{\Box}}_2^2}
\delta^{12}(z)| q^+_a(1) q^{+a}(2) q^+_b(2)q^{+b}(1) $$ $$ =\int
d\zeta^{(-4)}\frac{du_1du_2}{(u^+_1u^+_2)^2}\frac{({\cal D
}^+_2)^4}{{\stackrel{\frown}{\Box}}_2^2}\delta^{12}(z) D^{++}_1
q^-_a(1)q^{+a}(2)q^+_b(2)q^{+b}(1) $$ $$ =\int
d\zeta^{(-4)}du_1du_2 [D^{--}_1\delta^{(2,-2)}(u_1u_2)]
\frac{({\cal D
}^+_2)^4}{{\stackrel{\frown}{\Box}}_2^2}\delta^{12}(z)q^-_a(1)q^{+a}(2)q^+_b(2)q^{+b}(1)
$$ Writing $D^{--}_1\delta^{(-2,2)}(u_1u_2) =D^{--}_2
\delta^{(0,0)}(u_1u_2)$ and integrating over $u_2$ one finally  gets
$$ i\Gamma_4= \frac{1}{2}\int d\zeta^{(-4)}du (-4
q^-q^+)^2 \frac{({\cal D
}^+)^4}{{\stackrel{\frown}{\Box}}^2}\delta^{12}(z)|. $$
Here we have used the restrictions (\ref{rest}).

Analysis of a general term ${\Gamma}_{2n}$ is carried out
analogously. First of all we transform all analytic subspace
integrals into ones over the  full superspace taking the factors
$(D^+_k)^4(D^+_{k+1})^4$ from the hypermultiplet propagators. Then
we integrate over the sets of Grassmann and space-time coordinates
using the corresponding delta-functions in the integrand. It
leads to
\begin{equation}\label{3}
\int d^{12}z du_1 ... du_{2n} \frac{\delta^{(-2,2)}(u_2,u_3)
\delta^{(-2,2)}(u_4,u_5) ...
\delta^{(-2,2)}(u_{2n},u_1)}{(u^+_1u^+_2)^3(u^+_3u^+_4)^3 ...
(u^+_{2n-1}u^+_{2n})^3}
\end{equation}
$$\times \frac{({\cal D}^+_2)^4 ({\cal D}^+_4)^4 ... ({\cal
D}^+_{2n})^4}{{\stackrel{\frown}{\Box}}_{1}
{\stackrel{\frown}{\Box}}_{2} ... {\stackrel{\frown}{\Box}}_{2n}}
\{\delta^{12}(z-z')| q^+_a(u_1)q^{+a}(u_2)q^+_b(u_3) ...
q^+_c(u_{2n-1}) q^{+c}(u_{2n})\}. $$
Then we integrate over
$u_2, u_4, ..., u_{2n}$ using the harmonic delta functions and
obtain
\begin{equation}\label{6}
\int \frac{du_1 du_3 ...
du_{2n-1}}{(u^+_1u^+_3)^3(u^+_3u^+_5)^3...(u^+_{2n-1}u^+_1)^3}
\frac{({\cal D}^+_1)^4 ({\cal D}^+_3)^4...({\cal
D}^+_{2n-1})^4}{{\stackrel{\frown}{\Box}}_1^2
{\stackrel{\frown}{\Box}}_3^2...{\stackrel{\frown}{\Box}}_{2n-1}^2}
\end{equation}
$$\times \{\delta^{12}(z-z')|
q^+_a(u_1)q^{+a}(u_3)q^+_b(u_3)q^{+b}(u_5)...
q^+_c(u_{2n-1})q^{+c}(u_1)\}. $$ After relabelling the
indices  $c \rightarrow a, a \rightarrow b, ...;\quad 3
\rightarrow 2,..., (2n-1) \rightarrow n$ we get the expression
\begin{equation}\label{7}
i \Gamma_{2n}=\frac{4(-1)^n2^n}{n}\int d^{12}z du_1 ... du_n
\frac{({\cal D}^+_1)^4 ({\cal D}^+_2)^4 ... ({\cal
D}^+_n)^4}{(u^+_1u^+_2)^3 (u^+_2u^+_3)^3 ... (u^+_nu^+_1)^3}
\frac{1}{{\stackrel{\frown}{\Box}}_1^2
{\stackrel{\frown}{\Box}}_2^2 ...{\stackrel{\frown}{\Box}}_n^2}
\end{equation}
$$\times \{\delta^{12}(z-z')|_{z=z'}
q^{+a}(u_1)q^{+}_b(u_1)q^{+b}(u_2)q^{+}_c(u_2)... q^{+}_a(u_n)\}.
$$
To simplify the expression (\ref{7}) we represent $q^+_b
(u_1)$ as  $q^+_b (u_1)=D^{++}_1 q^-_b (u_1)$ (since $q^+$  sits
on its mass shell). We note that when acting on
${\stackrel{\frown}{\Box}}$, $D^{++}$ gives rise to the structures like
$({\cal D}^+)^5=0.$ Now we integrate by parts and  remove the
harmonic derivative on the harmonic distributions
\begin{equation}\label{8}
-D^{++}_1 \frac{1}{(u^+_1 u^+_2)^3(u^+_n u^+_1)^3} = \frac{1}{2}
\{(D^{--}_1)^2 \delta^{(3,-3)} (u_1,u_2) \frac{1}{(u^+_1 u^+_n)^3}
+ (2 \leftrightarrow n)\}
\end{equation}
and use the identity \cite{gikos}, \cite{gios}
\begin{equation}\label{9}
(D^{--}_1)^2 \delta^{(3,-3)}(1|2) = (D^{--}_2)^2 \delta^{(-1,1)}
(1|2).
\end{equation}
Then we take the factor$(D^{--}_2)^2$ off the harmonic delta
function. It is easy to see that this factor can give a
non-vanishing result only when acts on $({\cal D}^+(u_2))^4$. We get
\begin{equation}\label{10}
\int d u_1 ... (-\frac{1}{2}) \delta^{(-1,1)}(1|2) \frac{({\cal
D}^+_1)^4 (D^{--}_2)^2 ({\cal D}^+_2)^4 ... ({\cal
D}^+_n)^4}{(u^+_2 u^+_3)^3 ...(u^+_n u^+_1)^3}
\frac{1}{{\stackrel{\frown}{\Box}}^2_1
{\stackrel{\frown}{\Box}}^2_2 ...
{\stackrel{\frown}{\Box}}^2_n}\end{equation} $$\times
\{\delta^{12}(z)| q^{+a}(u_1) q^-_b(u_1)q^{+b}(u_2) ...
q^+_a(u_n)\} + (2 \leftrightarrow n).$$ We replace $u_2
\leftrightarrow u_n$ in the second term here, after that it
becomes identical to the first one. Doing integral over $u_1$ we
obtain the expression $(-1) (-2 {\stackrel{\frown}{\Box}}_2)
\frac{1}{{\stackrel{\frown}{\Box}}^4_2}$. At the second step, we
repeat the above procedure for $q^+_c(u_2)$ , i.e represent it in the
form  $q^+_c (u_2) = D^{++}_2 q^-_c$ and integrate by parts with
respect to ${\cal D}^{++}_2$. Performing the same manipulations as
above we obtain the factor $(-1)^2 (-2
{\stackrel{\frown}{\Box}}_3)^2
\frac{1}{{\stackrel{\frown}{\Box}}^6_3}.$

After $n-4$ analogous steps $q^+_d(u_3) = D^{++}_3 q^-_d(u_3)$
we reduce the harmonic integral in expression (\ref{7}) to
that over three sets of harmonic:
\begin{equation}\label{11}
\frac{({\cal D}^+_u)^4 ({\cal D}^+_{u_{n-1}})^4 ({\cal
D}^+_{u_n})^4}{{\stackrel{\frown}{\Box}}_u^{2n-4}
{\stackrel{\frown}{\Box}}_{u_{n-1}}^2
{\stackrel{\frown}{\Box}}_{u_n}^2} \frac{(2
{\stackrel{\frown}{\Box}})^{n-3}}{(u^+ u^+_{n-1})^3 (u^+_{n-1}
u^+_n)^3 (u^+_n u^+)^3} \{\delta^{12}(z-z')|\end{equation} $$
\times q^{+a}(u)...
q^+_c(u)q^{+c}(u_{n-1})q^+_d(u_{n-1})q^{+d}(u_n) q^+_a(u_n)\} $$
On last step we write $q^+_c (u) = D^{++}_u q^-_c$ and remove the
$D^{++}_u$ on the harmonic factor. Repeating the same
manipulations we perform the $u_{n-1}$-integration and get the
expression
\begin{equation}\label{12}
i \Gamma_{2n} = -\frac{(-2)^n 2^n}{n}\int d^{12}z d u du_1
\frac{({\cal D}^+_u)^4({\cal
D}^+_{u_1})^4}{{\stackrel{\frown}{\Box}}^n_u
{\stackrel{\frown}{\Box}}^2_{u_1}}\frac{1}{(u^+ u^+_1)^6}
\{\delta^{12}(z-z')| q^{+a}(u)\end{equation} $$\times
q^-_b(u)q^{+b}(u)... q^+_c(u)q^{+c}(u_1)q^+_a(u_1)\}.$$

Now ones return to analytic subspace applying the identities
(\ref{D}) $$ \int d\zeta^{(-4)}\frac{d u du_1}{(u^+
u^+_1)^2}\frac{({\cal
D}^+_u)^4}{{\stackrel{\frown}{\Box}}^n_u}\delta^{12}(z-z')|q^{+a}(u)q^-_b(u)q^{+b}(u)...
q^+_c(u)q^{+c}(u_1)q^+_a(u_1). $$ Then we use the identities
$q^+(u_1)=(u^+_1u^-)q^+(u)-(u^+_1u^+)q^-$, $q^-_a q^{-a}=0$
leading to the factor $(u^+_1 u^+_2)^2.$ We get finally
\begin{equation}\label{G2n}
i\Gamma_{2n}= \frac{1}{n}\int d\zeta^{(-4)}du \frac{({\cal
D}^+)^4}{{\stackrel{\frown}{\Box}}^n}\delta^{12}(z-z')|(-4q^-q^+)^n.
\end{equation}
Here we again have used the restrictions (\ref{rest}).

Now sum up all contributions ${\Gamma}_{2n}$ (\ref{G2n}).  The
result is given in terms of functional determinant of special
differential operator:
\begin{equation}\label{finally}
i\Gamma =-\int d\zeta^{(-4)}du \ln(1 +\frac{4q^-_a
q^{+a}}{{\stackrel{\frown}{\Box}}})({\cal
D}^+)^4\delta^{12}(z-z')|_{z'=z}
\end{equation}
$$ =-\int d\zeta^{(-4)}du \ln({{\stackrel{\frown}{\Box}}} +{4q^-_a
q^{+a}})({\cal D}^+)^4\delta^{12}(z-z')|_{z'=z} $$$$+ \int
d\zeta^{(-4)}du \ln({{\stackrel{\frown}{\Box}}})({\cal
D}^+)^4\delta^{12}(z-z')|_{z'=z}. $$
It is evident that the second
term in (\ref{finally}) is (up to the sign) the representation of the
one-loop effective action for ${\cal N}=4$ SYM theory in sector
the ${\cal N}=2$ vector multiplet \cite{2}, \cite{1b}. We see the
${\Gamma}$ in (\ref{finally}) vanishes when the hypermultiplets
vanish. We point out that ${\Gamma}$ (\ref{finally}) is only a
part of full effective action (\ref{log}) which essentially
depends on hypermultiplets. The full effective action contains
also hypermultiplet independent part ${\Gamma}(V^{++})$.

We would like to emphasize that in spite of the formal presence of
$q^{-}_{a}$ in (\ref{finally}), the integrand in (\ref{finally})
is an analytic superfield in framework of the low-energy
approximation (see the restrictions (\ref{rest} in the
hypermultiplet sector). Indeed, the combination $q^{-}_{a}q^{+a}$
is harmonic-independent \cite{1} and is proportional to
$q^{ia}q_{ia}$ where $q^{ia}$ is (constrained) superfield on
general ${\cal N}=2$ superspace. This quantity should be treated
as a constant if we consider only leading low-energy approximation
for hypermultiplet dependence (see the restrictions for background
hypermultiplet (\ref{rest}), which we use in the paper).
Such a  situation is completely consistent with a general picture
of finding the effective action as a series in background field
derivatives. If we look for the terms in effective action without
derivatives it is sufficient to consider all possible derivatives
of background fields to be equal zero. Therefore the Eq.
(\ref{finally}) is completely correct under approximation what was
formulated in the paper from the very beginning \footnote{We point
out once more that we are considering only the leading derivative
independent hypermultiplet contribution to the low-energy
effective action. Of course, such an approximation violates the
invariance under the hidden ${\cal N}=2$ supersymmetry. It would
be extremely interesting and useful to develop a systematic
procedure to derive  the derivative expansion of the effective
action in hypermultiplet sector, However this problem is beyond
the scope of this paper.}.

Thus,  using the ${\cal N}=2$ harmonic superspace formulation of
${\cal N}=4$ SYM theory and techniques of harmonic supergraphs
we obtained the representation of the one-loop effective action in
hypermultiplet sector (\ref{finally}). This representation is free
of harmonic singularities and, as we will see,  admits a
straightforward evaluation with help of ${\cal N}=2$ superfield
heat kernel method. We will show in the next section that the general
expression (\ref{finally}) allows us to obtain the exact
proper-time representation of the effective action and its expansion in
covariant spinor derivatives of
the ${\cal N}=2$ superfield Abelian strengths ${\cal W}, \bar{\cal
W}$ corresponding to the constant space-time superstrength background
\begin{equation}\label{const}
{\cal W}|_{\theta=0}=\mbox{const}, D^{\pm}_{\alpha}{\cal
W}|_{\theta=0}=\mbox{const}, \bar{D}^{\pm}_{\dot\alpha}\bar{\cal
W}|_{\theta=0}=\mbox{const}, D^-_\alpha D^+_\beta {\cal
W}|_{\theta=0}=\mbox{const}, \bar{D}^-_{\dot\alpha}
\bar{D}^+_{\dot\beta} \bar{\cal W}|_{\theta=0}=\mbox{const}
\end{equation}
and the on-shell background hypermultiplet $q^{\pm a} =
q^{ia}u^{\pm}_{i}$ where $q^{ia}$ does not depend of
harmonics \cite{1} and can be taken to be a constant.

The Eq. (\ref{finally}) immediately leads to one of the main
results concerning the hypermultiplet dependence of one-loop
effective action: the hypermultiplet enters to the effective
action in the combination ${\bar{\cal W}}{\cal W} +
2q^{-}_{a}q^{+a}$ which is invariant under the $R$-symmetry of ${\cal
N}=4$ supersymmetry  for constant ${\cal W}$ and on-shell
hypermultiplet \cite{1}. To see that we consider the  expression
${\stackrel{\frown}{\Box}} + 4q^{-}_{a}q^{+a}$ and use the
on-shell form of ${{\stackrel{\frown}{\Box}}}$ (\ref{07}). Then
${\stackrel{\frown}{\Box}} + 4q^{-}_{a}q^{+a} = \frac{1}{2}{\cal
D}^{\alpha\dot\alpha}{\cal D}_{\alpha\dot\alpha}
 -\frac{i}{\sqrt{2}}((D^{+\alpha}{\cal
W}){\cal D}^-_{\alpha} + (\bar{D}^+_{\dot\alpha}\bar{\cal
W})\bar{\cal D}^{- \dot\alpha})+2{\cal W}\bar{\cal W} +
4q^{-}_{a}q^{+a}$. We see that this operator contain just the
above combination (two last terms without derivatives) and just
this combination will be on of the arguments of hypermultiplet
dependent effective action (36) in the low-energy
approximation\footnote{It is worth pointing out that such an
approximation violates $R$-symmetry of initial classical model.
Problem of constructing the general derivative expansion of
effective action containing the hypermultiplet spinor derivatives
and preserving the $R$-symmetry is open at present. We consider
this work as a first step towards this generic problem.}.


\section{The proper-time representation of the effective action}


Relation (\ref{finally}) possesses remarkable features. First, we
started with the model of two interacting fields $V^{++}$, $q^{+}$
and resummed the supergraphs by such a way that the effective
action is expressed in terms of a differential operator acting
only in the sector of the vector multiplet. The whole dependence on
the hypermultiplets is included in this operator. Second, the
effective action is written as an integral over the analytic
subspace of the harmonic superspace. It is co-ordinated with the
classical action of the theory which is also written as an
integral over the analytic subspace. Third, the relation
(\ref{finally}) has the form $\mbox{Tr}\ln\hat{A}$ with the
operator $\hat{A} = {\stackrel{\frown}{\Box}} + 4q^{-}_{a}q^{+a}$
acting on the analytic superfields. We emphasize that the above
simple form of the one-loop effective action is not evident from
the very beginning, it is the result of resummation of the
infinite sequence of the one-loop harmonic supergraphs with
arbitrary number hypermultiplet external legs. The form of the
effective action (\ref{finally}) is basic for using the
proper-time representation:
\begin{equation}\label{heat1}
\Gamma=i \int d\zeta^{(-4)} du \int^{\infty}_0
\frac{ds}{s}e^{-s({{\stackrel{\frown}{\Box}}} +{4q^-_a
q^{+a}})}({\cal D}^+)^4\delta^{12}(z-z')|_{z'=z}$$$$= i
\int^{\infty}_0 \frac{ds}{s}\mbox{Tr}\{K(s)e^{-s(4q^-_a
q^{+a})}\}~,
\end{equation}
Here $K(s)$ is a superfield heat kernel, the operation $\mbox{Tr}$
means the functional trace in the analytic subspace of the
harmonic superspace $\mbox{Tr}K(s)= \mbox{tr} \int
d\zeta^{(-4)}K(\zeta,\zeta|s)$, where $\mbox{tr}$ denotes the
trace over the discrete indices. As a result, the problem of
finding the one-loop effective action is reduced to evaluation of
the kernel $K(s) = e^{-s{\stackrel{\frown}{\Box}}}$. Further it is
convenient to deal with ${\stackrel{\frown}{\Box}}$ written by a
definition as\footnote{Note that the replacements ${\cal W}
\rightarrow \frac{{\cal W}}{-\sqrt{2}}$, $\bar{\cal W} \rightarrow
\frac{\bar{\cal W}}{-\sqrt{2}}$ reduces the expression (\ref{7a})
to (\ref{07}).}
\begin{equation}\label{7a}{\stackrel{\frown}{\Box}}
=\frac{1}{2}{\cal D}^{\dot\alpha \alpha} {\cal D}_{\alpha
\dot\alpha} + \frac{i}{2} (D^{+ \alpha}{\cal W}){\cal
D}^-_{\alpha} +\frac{i}{2}(\bar{D}^+_{\dot\alpha} \bar{\cal W})
\bar{\cal D}^{- \dot\alpha} + {\cal W} \bar{\cal W}.
\end{equation}

For the effective action calculation (\ref{heat1}) we will use the
techniques of symbols of operators in the analytic subspace of
${\cal N}=2$ harmonic superspace \footnote{Applications of
techniques of symbols of operators for calculating the effective
action in ${\cal N}=1$ superspace are given in \cite{pb}.}. We
begin with the Fourier representation of the delta-function in
${\cal N}=2$ superspace
\begin{equation}
\delta^{12}(z-z') =\int \frac{d^4 p}{(2 \pi)^4} \int d^4 \psi^+
d^4 {\psi}^-e^{i p_m (x-x')^m}
\end{equation}
$$
\times e^{(\theta-\theta')^{+ \alpha}
\psi^-_{\alpha}+\bar{\psi}^-_{\dot\alpha}(\bar{\theta}
-\bar{\theta}')^{+ \dot\alpha}}e^{(\theta-\theta')^{-\alpha}
\psi^+_{\alpha}+\bar{\psi}^+_{\dot\alpha}(\bar{\theta}
-\bar{\theta}')^{-\dot\alpha}}~,
$$
where $p_{M}=\{p_m,\psi^{\pm}_\alpha,\bar{\psi}^{\pm}_{\dot\alpha}\}$
is a cotangent supervector at a superspace point $z$ and $d^4\psi^+
=\\
\frac{1}{16} d^2 \psi^+ d^2 \bar{\psi}^+.$ As a result, the heat
kernel at coincident points takes the form
\begin{equation}\label{X}
K(\zeta,\zeta|s)=\int \frac{d^{4}p}{(2\pi)^4}d^8 \psi \exp\{-s (\frac{1}{2}({\cal D}+ip)^{\dot\alpha \alpha} ({\cal D}+ip)_{\alpha
\dot\alpha} + \frac{i}{2} (D^{+ \alpha}{\cal W})({\cal
D}+\psi)^-_{\alpha}
\end{equation}
$$ +\frac{i}{2}(\bar{D}^+_{\dot\alpha} \bar{\cal W}) (\bar{\cal
D}+\bar{\psi})^{- \dot\alpha} + {\cal W} \bar{\cal W})\} ({\cal
D}^+ +\psi^+)^{4}\times{\bf 1}. $$
Then, we have to act by the
operators in the exponential to the right on the unity. After that,
all differential operators can act only on ${\cal W}, \bar{\cal
W}$. Therefore, the final result is expressed in terms of
strengths ${\cal W}, \bar{\cal W}$ and their spinor derivatives.
This procedure can be organized, for example, as follows. We
extract in (\ref{X}) the exponential of the leading symbol
$\frac{1}{2}p^{\alpha\dot\alpha}p_{\alpha\dot\alpha}$ of the
operator (\ref{7a}) and expand the rest exponential in a power
series in covariant derivatives. Then we move all the derivatives
to the right, commuting or anticommuting them with the
coefficients of the operator (\ref{7a}), act on unity and kill
them. Last step is Gaussian integration over momenta $p$ and
trivial integration over odd variables $\psi$. The result will be
a series in covariant derivatives of the strengths.  However, this
way of consideration demands a lot of straightforward calculations
to get a final result in manifest covariant form.

A method generating the manifest supersymmetric asymptotical
expansion of heat kernel in ${\cal N}=1$ superspace has been
developed in \cite{garg}, \cite{pb}, \cite{bbp}. We generalize
this method for the heat kernel in ${\cal N}=2$ harmonic
superspace. In each superspace point we introduce a tangent space
forming the normal-coordinate system and a fiber frame obtained by
a parallel transport from the base point. The pseudodifferential
operators can be reexpressed in this local representation of the
vector bundle. We consider the heat kernel using these operators
and derive an algorithm for the asymptotic expansions of the heat
kernel.

Let us introduce the following notions
\begin{equation}\label{salg1}
A^{+\alpha}=\frac{i}{2}[D^{+\alpha}, {\cal W}],\;
\bar{A}^{+\dot\alpha}=-\frac{i}{2}[\bar{D}^{+\dot\alpha},
\bar{\cal W}], \; A^{-\alpha}=[D^{-\alpha},{\cal W}],\;
\bar{A}^{-\dot\alpha}=[\bar{D}^{-\dot\alpha},\bar{\cal W}],
\end{equation}
$$ \{D^-_\alpha,
A^+_\beta\}=N_{\alpha\beta}=N_{\beta\alpha}=\frac{i}{2}D^-_\alpha
D^+_\beta {\cal W},
\;\{\bar{D}^-_{\dot\alpha},\bar{A}^+_{\dot\beta}\}=\bar{N}_{\dot\alpha
\dot\beta}=\bar{N}_{\dot\beta\dot\alpha}=-\frac{i}{2}\bar{D}^-_{\dot\alpha}
\bar{D}^+_{\dot\beta}\bar{\cal W}. $$ In this terms the algebra of
covariant derivatives (\ref{algebra}) takes the form:
\begin{equation}\label{salg2}
[{\cal D}_{\alpha\dot\alpha}, x^{\beta\dot\beta}]=2
\delta^{\beta\dot\beta}_{\alpha\dot\alpha}, \quad \{{\cal
D}^-_\alpha, \theta^{+\beta}\}=\delta^\beta_\alpha, \quad
\{\bar{\cal D}^-_{\dot\alpha},
\bar{\theta}^{+\dot\beta}\}=\delta^{\dot\beta}_{\dot\alpha}~,
\end{equation}
\begin{equation}\label{salg3}
[{\cal D}^{\dot\alpha \alpha},{\cal
D}_{\beta\dot\beta}]=-(\delta^{\dot\alpha}_{\dot\beta}N^\alpha_\beta
+ \delta^{\alpha}_\beta \bar{N}^{\dot\alpha}_{\dot\beta})
=-F^{\alpha \dot\alpha}_{\beta\dot\beta}~,
\end{equation}
$$ [{\cal D}^-_\alpha ,{\cal D}_{\beta
\dot\beta}]=\varepsilon_{\alpha\beta}\bar{A}^-_{\dot\beta},\quad
[\bar{\cal D}^-_{\dot\alpha}, {\cal
D}_{\beta\dot\beta}]=\varepsilon_{\dot\alpha \dot\beta}A^-_\beta~.
$$ $$ {\cal D}_m A^{\pm}_{(\alpha,\dot\alpha)}={\cal
D}^{\pm}_{(\delta, \dot\delta)}N_{\alpha\beta}={\cal
D}^{\pm}_{(\delta, \dot\delta)}\bar{N}_{\dot\alpha\dot\beta}={\cal
D}_m N_{\alpha\beta}={\cal D}_m \bar{N}_{\dot\alpha\dot\beta}=0.
$$ We see that the  set of covariant derivatives together with the
on-shell background superfields ${\cal W}, \bar{\cal W}$,
corresponding to the constant space-time configurations
(\ref{const}), and their low-order derivatives generates a finite
dimensional Lie superalgebra (\ref{salg1})-(\ref{salg3}).

At the next step, we lift the shifted operators $X_m={\cal D}_m
+ip_m, X^{\pm}_{(\alpha,\dot\alpha)}={\cal
D}^{\pm}_{(\alpha,\dot\alpha)}+\psi^{\pm}_{(\alpha,\dot\alpha)}$
satisfying the same algebra (\ref{salg1})-(\ref{salg3}) in the
tangent space by the "exponential map" $X (p_M,\partial/\partial
p_M)= U^{-1} X_M U$\footnote{Main property of this transformation
is to eliminate the operators ${\cal D}_m, {\cal D}^{+}_{\alpha},
{\cal D}^{-}_{\dot{\alpha}}$ in $X_{M}$. Further we use the same
notation $X_{M}$ for both transformed and initial quantities.}
 with
$$
U = e^{-\partial^{-\alpha}_\psi {\cal D}^+_{\alpha} -\bar{\cal
D}^+_{\dot\alpha} \bar{\partial}^{-\dot\alpha}_\psi} \cdot
e^{2\theta^{- \alpha} p_{\alpha \dot\alpha}\bar{\partial}^{+
\dot\alpha}_\psi - 2 \partial^{+ \alpha}_\psi p_{\alpha
\dot\alpha} \bar{\theta}^{- \dot\alpha}} \cdot e^{\partial^{+
\alpha}_\psi{\cal D}^-_{\alpha} + \bar{\cal D}^-_{\dot\alpha}
\bar{\partial}^{+\dot\alpha}_\psi}\cdot e^{-\frac{i}{2}
\partial^{\dot\alpha \alpha}_p {\cal D}_{\alpha \dot\alpha}}~,
$$ where the role of a tangent vectors forming a normal-coordinate
frame plays a set of derivatives $$
\partial^{\mp\alpha}_\psi \equiv \frac{\partial}{\partial
\psi^{\pm}_{\alpha}}, \quad \bar{\partial}^{\mp\dot\alpha}_\psi
\equiv \frac{\partial}{\partial \bar{\psi}^{\pm}_{\dot\alpha}},
\quad
\partial ^{\alpha\dot\alpha}_p \equiv \frac{\partial}{\partial p_{\alpha
\dot\alpha}} \quad [\partial^M, p_N\}=\delta^M_N. $$ The action of
the operator $U$ on the shifted operators $X_M$ is given by
following expressions $$ U^{-1}(X^+_{\alpha}) U = \psi^+_{\alpha},
\quad U^{-1}(\bar{X}^+_{\dot\alpha})U = -
\bar{\psi}^+_{\dot\alpha},$$ $$ U^{-1}(X^-_{\alpha}) U =
-\psi^-_{\alpha} +2\bar{\partial}^{- \dot\alpha}_\psi p_{\alpha
\dot\alpha}+ {\cal O}(\partial^-_\psi,
\partial_p ),\quad U^{-1}(\bar{X}^-_{\dot\alpha}) U =
\bar{\psi}^-_{\dot\alpha} +2\partial^{- \alpha}_\psi p_{\alpha
\dot\alpha}+ {\cal O}(\partial^-_\psi,\partial_p), $$ $$ \{
X^+_{\alpha}, \bar{X}^-_{\dot\alpha} \}= 2 X_{\alpha \dot\alpha}=
-\{ X^-_{\alpha}, \bar{X}^+_{\dot\alpha}\},$$
\begin{equation}\label{U}
X_{\alpha \dot\alpha} =i p_{\alpha \dot\alpha} +\partial^+_{\psi
\alpha} (\bar{D}^-_{\dot\alpha} \bar{\cal W})
-\bar{\partial}^+_{\psi\dot\alpha}(D^-_\alpha{\cal W})
-\frac{1}{8} \{\partial^{\dot\beta}_{p \alpha}
(\bar{D}^+_{\dot\beta} \bar{D}^-_{\dot\alpha}\bar{\cal W})
+\partial^\beta_{p \dot\alpha} (D^-_\beta D^+_\alpha {\cal W}\}+
{\cal O}(\partial^-_\psi, \partial_p).
\end{equation}
The map of the function on the superspace in the point $z$ into
the tangent superspace is given by
\begin{equation}\label{Ua}
{\cal W} \rightarrow {\cal W} -\partial^{+\alpha}_\psi (D^-_\alpha
{\cal W})+{\cal O}(\partial^-_\psi, \partial_p) ,\quad \bar{\cal
W} \rightarrow \bar{\cal W} +\bar{\partial}^{+
\dot\alpha}_\psi(\bar{D}^-_{\dot\alpha}\bar{\cal W})+{\cal
O}(\partial^-_\psi, \partial_p),
\end{equation}
$$ D^+_{\alpha}{\cal W} \rightarrow D^+_{\alpha}{\cal W}
-\partial^{+\beta}_\psi (D^-_{\beta}D^+_{\alpha}{\cal W})+{\cal
O}(\partial^-_\psi, \partial_p),$$$$
\bar{D}^+_{\dot\alpha}\bar{\cal W}\rightarrow
\bar{D}^+_{\dot\alpha}\bar{\cal W} +
\bar{\partial}^{+\dot\beta}_\psi
(\bar{D}^-_{\dot\beta}\bar{D}^+_{\dot\alpha} \bar{\cal W})+{\cal
O}(\partial^-_\psi, \partial_p), $$ $$ D^-_\alpha {\cal
W}\rightarrow D^-_\alpha {\cal W}+{\cal O}(\partial^-_\psi,
\partial_p), \quad \bar{D}^-_{\dot\alpha} \bar{\cal W} \rightarrow
\bar{D}^-_{\dot\alpha} \bar{\cal W}+{\cal O}(\partial^-_\psi,
\partial_p). $$ We point out that in this representation the
operators $X$, superfields and their derivatives satisfy the same
algebra (\ref{salg2}), (\ref{salg3}). Generally speaking, if the
background superfields are arbitrary, all above quantities  are
presented by an infinite series over $\partial_p$ and a finite
series over Grassmann derivatives $\partial_\psi$ with the
coefficients in a fixed point $z^A$. But for the considered
background (on-shell and space-time constant background fields)
the representation (\ref{U}, \ref{Ua}) is exact.

The actual calculation of the effective action (\ref{heat1}) with
the kernel (\ref{X}) is based on the following observation. The
operator $e^{-s{\stackrel{\frown}{\Box}_{U}}}$, where the operator
${\stackrel{\frown}{\Box}_{U}}$ is given in terms of shifted
variables $X$ (\ref{U}), can be considered as an evolution
operator for a Bose-Fermi quantum system with a Hamiltonian
$\hat{H}={\stackrel{\frown}{\Box}_{U}}$. Eqs. (\ref{7a}),
(\ref{U}) show that the Hamiltonian $\hat{H}$ is a quadratic form
in the operators $p, \partial_p, \psi, \partial_{\psi}$ with
constant coefficients (due to background under consideration).
Therefore, calculation of $TrK(s)=\int
d{\zeta}^{-4}K(\zeta,\zeta|s)$ with $K(\zeta,\zeta|s)$ given by
(\ref{X}), is an exactly solvable problem.

Let us return to Eq. (\ref{X}) where all operators and fields
along with their derivatives are written in the representation
(\ref{U}). This is equivalent to an extension of the heat kernel
to the tangent bundle on the superspace point $z$ where the
coordinate $z$ is considered as a constant parameter. According to
(\ref{X}) the evolution operator should act on the unity. It is
evident that a result of such action is obtained if we recommute
all derivatives with $p$ and $\psi$ to the right and omit them.
This procedure can be realized on the base of the
Baker-Campbell-Hausdorff formula corresponding to the algebra
(\ref{salg2}), (\ref{salg3}) (see Appendix A). The result we
obtain is called a symbol of the evolution operator. This symbol
has to be integrated over bosonic $p$ and fermionic $\psi$
variables, which leads us to the heat kernel (\ref{X}). Next useful
observation is based on the fact that the exponential in the
evolution operator contains only ${\cal D}^- \sim \psi^-$. All
${\cal D}^+ \sim \psi^+$ are in the pre-exponential factor $({\cal
D}^{+4})$ and saturate the integral over $d^4 \psi^{+}$. Therefore
we must omit all ${\cal O}(\partial^-_\psi)$ in the  operator
$\hat{H}$ that gives us a more simple expression for (\ref{U},
\ref{Ua}). In addition, to do the Berezin integral over $\psi^-$
sector we have to extract a "projector" $(\psi^-)^4$ from the
exponential under consideration.

This program can be effectively realized if we present the
exponent $K(s) = e^{-s{\stackrel{\frown}{\Box}}}$ as a products of
several operator exponents. Such a construction allows to overcome
the difficulties which arose in previous attempts of computing the
effective action of ${\cal N}=2$ SYM theories directly in ${\cal
N}=2$ superspace. We write the operator $K(s) =
e^{-s{\stackrel{\frown}{\Box}}}$ in the form
\begin{equation}\label{h3}
K(s)=\exp (-s\{A^+{\cal D}^- + \bar{A}^+\bar{\cal D}^-
+\frac{1}{2}{\cal D}^{\dot\alpha\alpha}{\cal D}_{\alpha\dot\alpha}
+ {\cal W}\bar{\cal W}\})
\end{equation}
$$ =\exp\{-f_{\alpha\dot\alpha}(s){\cal
D}^{\dot\alpha\alpha}\}\exp\{-s\frac{1}{2}{\cal
D}^{\dot\alpha\alpha}{\cal D}_{\dot\alpha\alpha}\}
\exp\{-\Omega(s)\}\exp\{-s(A^{+\alpha}{\cal D}^-_{\alpha}
+\bar{A}^{+\dot\alpha}\bar{\cal D}^-_{\dot\alpha})\}~. $$ with
some unknown coefficients in the right hand side. These
coefficients can be found directly, i.e. using the
Baker-Campbell-Hausdorff formula (representation of the
Baker-Campbell-Hausdorff formula we use is given in Appendix A),
and by solution to the system of a differential equation on the
coefficients. Both ways lead to the same results. To find the
mentioned system of equations we consider $(\frac{d}{ds}K) K^{-1}$
and substitute for $K$ first and second lines in (\ref{h3})
subsequently.

Equations for the functions $f^{\dot\alpha\alpha}(s)$ have the form
\begin{equation}\label{h4}
\dot{f}_{\alpha\dot\alpha}(s)=
-f_{\beta\dot\beta}F^{\dot\beta\beta}_{\dot\alpha\alpha} -
A^{+\beta}(D^-_\beta f_{\alpha\dot\alpha})
-\bar{A}^{+\dot\beta}(\bar{D}^-_{\dot\beta} f_{\alpha\dot\alpha})
\end{equation}
$$+A^+_\beta \bar{A}^-_{\dot\beta}(\int^s_0 d\tau e^{\tau F}
)^{\dot\beta\beta}_{\dot\alpha\alpha} + \bar{A}^+_{\dot\beta}
{A}^-_{\beta}(\int^s_0 d\tau e^{\tau F}
)^{\dot\beta\beta}_{\dot\alpha\alpha}~,
$$
Analogously, equation for the function $\Omega$ is
\begin{equation}\label{h5}
\dot{\Omega}(s) -{\cal W}\bar{\cal
W}=-A^{+\alpha}(D^-_\alpha\Omega)
-\bar{A}^{+\dot\alpha}(\bar{D}^-_{\dot\alpha}\Omega) +A^+_\alpha
f^{\alpha\dot\alpha}\bar{A}^-_{\dot\alpha}
+\bar{A}^+_{\dot\alpha}f^{\dot\alpha\alpha}A^-_\alpha
\end{equation}
$$
-\frac{1}{2}A^+_\beta \bar{A}^-_{\dot\beta}(\int^s_0 d\tau
e^{-\tau F}
)^{\dot\beta\beta}_{\dot\alpha\alpha}F^{\dot\alpha\alpha}_{\dot\rho\rho}
f^{\dot\rho\rho} -\frac{1}{2}\bar{A}^+_{\dot\beta}
{A}^-_{\beta}(\int^s_0 d\tau e^{-\tau F}
)^{\dot\beta\beta}_{\dot\alpha\alpha}F^{\dot\alpha\alpha}_{\dot\rho\rho}
f^{\dot\rho\rho}~.
$$
It is easy to show that the solution to equation (\ref{h4}) can be written as
\begin{equation}\label{h6}
f_{\alpha\dot\alpha}= -A^+_\delta{\cal
N}^{\delta\dot\delta}_{\alpha\dot\alpha}(s)\bar{A}^-_{\dot\delta}
-\bar{A}^+_{\dot\delta}\bar{\cal
N}^{\dot\delta\delta}_{\dot\alpha\alpha}(s)A^-_\delta,
\end{equation}
where the functions ${\cal N}(N,\bar{N}), \bar{\cal N}(N,\bar{N})$
are listed in the Appendix B. Solution of the equation (\ref{h5})
has the form
\begin{equation}\label{h8}
\Omega(s) = s{\cal W}\bar{\cal W} +A^{+\alpha}\Omega^-_\alpha (s)
+ \bar{A}^{+\dot\alpha}\bar{\Omega}^-_{\dot\alpha} (s)  +
(A^+)^2\Psi^{(-2)}(s) +(\bar{A}^+)^2\bar{\Psi}^{(-2)}(s)
\end{equation}
$$ +A^{+\alpha}\bar{A}^{+}_{\dot\alpha}\Psi^{
\dot\alpha(-2)}_{\alpha}(s). $$ We point out that this solution is
a finite order polynomials in powers of the Grassmannian elements
$A^{\pm}$, $\bar{A}^{\pm}$ (\ref{salg1}). The coefficients
$\Omega^-_\alpha (s)$, $\bar{\Omega}^-_{\dot\alpha} (s)$,
$\Psi^{(-2)} (s)$, $\bar{\Psi}^{(-2)} (s)$,
$\Psi^{\dot\alpha(-2)}_{\alpha} (s)$ are given in the Appendix B.

Now it is useful to write the last exponential in
(\ref{h3}) in the form
\begin{equation}\label{h14}
e^{-s(A^+{\cal D}^- +\bar{A}^+\bar{\cal D}^-)}=1 +
a^{+\alpha}{\cal D}^-_\alpha + \bar{a}^{+\dot\alpha}{\cal
D}^-_{\dot\alpha} +f^{+2}({\cal D}^-)^2 +\bar{f}^{+2}(\bar{\cal
D}^-)^2 +f^{+2\dot\alpha\alpha}{\cal D}^-_\alpha \bar{\cal
D}^-_{\dot\alpha}
\end{equation}
$$ +\bar{\Xi}^{+3\dot\alpha}\bar{\cal D}^-_{\dot\alpha}({\cal
D}^-)^2 +{\Xi}^{+3\alpha}{\cal D}^-_{\alpha}(\bar{\cal D}^-)^2 +
\Omega^{+4}({\cal D}^-)^2(\bar{\cal D}^-)^2~. $$ The coefficients
of this expansion can be found exactly and are given in the
Appendix B. We specially point out that $\Omega^{+4}(s) \sim
(A^{+})^4$. The integrand of the kernel (\ref{X}) is represented
as a product of the Schwinger type kernel $e^{-s\frac{1}{2}{\cal
D}^{\alpha\dot\alpha}{\cal D}_{\alpha\dot\alpha}}$ and terms of
expansion (\ref{h14}) in powers of ${\cal D}^{-}$ as well as of
expansion $e ^{-f_{\alpha\dot\alpha}{\cal D}^{\alpha\dot\alpha}},
\; e^{-\Omega (s)}$ in powers of Grassmann variables $A^+,
\bar{A}^+$. It leads to
\begin{equation}
K(s)=\int \frac{d^4 p}{(2\pi)^4} d^8\psi
e^{-s\frac{1}{2}X^{\alpha\dot\alpha}X_{\alpha\dot\alpha}}\{1 +
\frac{1}{2}f_{\alpha\dot\alpha}(s) X^{\alpha\dot\alpha}
f_{\beta\dot\beta}(s)X^{\beta\dot\beta}+...\}\end{equation}
$$\times\; e^{-\Omega(s)}\{1+
...+\Omega^{+4}(s)(\psi^-)^4\}(\psi^+)^4 \times {\bf 1} $$
This
very complicated expression can be significantly simplified using
remarkable properties of the Berezin integral. Only last term in
the last braces gives a nontrivial contribution. Since
$\Omega^{+4}$ $ \sim (A^+)^4$ and all $f_{\alpha\dot\alpha}$,
$\Omega$ are constructed from the elements $A^+, \bar{A}^+$ we
must omit in the above expression for $K(s)$ all this functions
except $e^{-s{\cal W}\bar{\cal W}}$. As a  result we have
\begin{equation}\label{fin5}
K(s)=\int \frac{d^4p}{(2\pi)^4} K_{Sch} (s)e^{-s{\cal W}\bar{\cal
W}} \Omega^{+4}(s)
\end{equation}
Last step of the consideration is computing the Schwinger type
kernel for the operator ${\Box}_{cov}(X_{m})=
\frac{1}{2}X^{\alpha\dot\alpha}X_{\alpha\dot\alpha},$ $K_{Sch}(s)
\equiv \int \frac{d^4p}{(2\pi)^4} \cdot e^{-s\Box_{cov}(X_{m})},$
where the operators $X_{\alpha\dot\alpha}$ are given in (\ref{U}).
Such a computation is standard now (see e.g. \cite{garg} for
details)\footnote{ We note that taking the operation ${\mbox tr}$
in the Lorentz indices is trivial due to the identity $$
N_\alpha^\beta N_\beta^\delta =-\frac{1}{4}D^-_\alpha D^{+\beta}
{\cal W} D^-_\beta D^{+\delta} {\cal W}=-\frac{1}{8}D^-_\alpha
D^{+\beta}D^-_\beta D^{+\delta} {\cal
W}^2=-\frac{1}{2}\delta^\delta_\alpha (D)^4 {\cal
W}^2=\delta^\delta_\alpha N^2$$}. We write down only the final
result
\begin{equation}\label{g9}
K_{Sch}(s)=\frac{i}{(4\pi s
)^2}\frac{s^2(N^2-\bar{N}^2)}{\cosh(sN)-\cosh(s\bar{N})}~.
\end{equation}
Here $N$ is given by
$N=\sqrt{-\frac{1}{2}D^4 {\cal W}^2}$.
It can be
expressed in terms of the two invariants of the Abelian vector
field ${\cal F}=\frac{1}{4}F^{mn}F_{mn}$ and
${\cal G}=\frac{1}{4}^{\star}F^{mn} F_{mn}$ as $N=\sqrt{2({\cal F} +i{\cal G})}.$
In the context of ${\cal N}=4$ SYM theory
this kernel has been found
in \cite{fra}, \cite{bkt} on the base of various approaches.
Here, we derived the kernel (\ref{g9}) completely in terms of
${\cal N}=2$ harmonic superspace.


\section{Effective action and its spinor covariant derivative expansion}


In the previous section we developed the proper-time techniques in
the harmonic superspace. Now we apply this techniques to
construct the effective action.

Effective action is written in the form (\ref{heat1}) where the
heat kernel at coincident points is given by (\ref{X}). We apply
the decomposition (\ref{fin5}), (\ref{g9}) for $K_{Sch}(s)$
(\ref{h22}) for $\Omega^{(+4)}$ and take into account that (i) Eq.
(\ref{X}) already contains $(\psi^{+})^4$ and hence we can use
immediately $\int d^4\psi^+ (\psi^+)^4=1$, (ii) to saturate the
integration over $\psi^-$ it is sufficient to keep in Eq.
(\ref{h14}) only the last term $\Omega^{+4}(\psi^-)^4$, (iii)
since $\Omega^{+4}$ $\sim (A^+)^4$  (\ref{h22}) we have to omit in
(\ref{h3}) all terms which dependent on $A^+$. It leads to the
effective action in the final form
\begin{equation}\label{N4}
\Gamma= \frac{1}{(4\pi)^2}\int d\zeta^{(-4)} du \int^{\infty}_0
\frac{ds}{s^3}e^{-s({2\cal W}\bar{\cal W} +4q^-_a
q^{+a})}\frac{s^2(N^2-\bar{N}^2)}{\cosh(sN)-\cosh(s\bar{N})}
\end{equation}
$$
\times\; \frac{1}{16}(D^+{\cal W})^2 (\bar{D}^+\bar{\cal W})^2
\frac{\cosh(sN)-1}{N^2} \frac{\cosh(s\bar{N})-1}{\bar{N}^2}
$$
One can show that the integrand in (\ref{N4}) can be expended in power series in
the quantities $s^2N^2, s^2\bar{N}^2$. After
change of proper time $s$ to $s'=s2{\cal
W}\bar{\cal W}$ we get the expansion in powers of $s'^2
\frac{4N^2}{4({\cal W}\bar{\cal W})^2}$ and their conjugate. Besides, we point out, since
the integrand of (\ref{N4}) is already $\sim (A^+)^4$ we can
change in  each term of expansion the quantities $N^2, \bar{N}^2$
by superconformal invariants $\Psi^2$ and $\bar{\Psi}^2$ \cite{bkt} expressing these quantities
from
\begin{equation}\label{scinv}
\bar{\Psi}^2=\frac{1}{\bar{\cal W}^2}D^4\ln {\cal W}= \frac{1}{2
\bar{\cal W}^2} \{\frac{N_\alpha^\beta N_\beta^\alpha}{{\cal W}^2}
+4\frac{A^{+\alpha} N_\alpha^\beta A^-_\beta}{{\cal W}^3}
+3\frac{(A^+)^2(A^-)^2}{{\cal W}^4}\}
\end{equation}
and its conjugate. After that, one can show that each term of the
expansion can be rewritten as an integral over general ${\cal
N}=2$ superspace.

The expression (\ref{N4}) at vanishing hypermultiplets has been
obtained in Refs. \cite{bkt}, \cite{2} by other methods.
Hypermultiplet dependent effective action was derived in
\cite{bbp} in terms ${\cal N}=1$ superfields, its transformation
to ${\cal N}=2$ harmonic form has been done in \cite{bbp} on the
base of a heuristic prescription how to reconstruct the effective
action, given in terms of ${\cal N}=1$ superfields, in manifest
${\cal N}=2$ supersymmetric form. Now we justify the prescription
used in \cite{bbp} and derive the hypermultiplet depended
effective action (\ref{N4}) completely in terms of harmonic
superspace.

Now we give a few first terms of the effective action expansion
(\ref{N4}) in a power series in $N^2, \bar{N}^2$ and compare them
with the results of ${\cal N}=1$ superspace calculations under the
special prescription about reconstruction of manifest ${\cal N}=2$
supersymmetric form of the effective written in terms of ${\cal
N}=1$ superfields. It was done in the work \cite{bbp} on the base
of a spinor covariant derivative expansion of the effective
action. We will see that spinor derivative expansion of (\ref{N4})
actually reproduces the expansion in \cite{bbp}. Each term of the
spinor covariant derivative expansion of the effective action
contains a definite power of the Abelian strength $F_{mn}$. Such
an expansion allows to extract an explicitly dependence of
$q^-q^+$. Since $N^2, \bar{N}^2$ includes the spinor covariant
derivatives of superstrengths (see (\ref{salg1}), this is just
expression in spinor covariant derivatives of ${\cal W}, \bar{\cal
W }$ . We use the expansions of $K_{Sch}(s)$ and $\Omega^{+4}(s)$.
It leads to $$ \Gamma=\frac{1}{(8\pi)^2}\int
d\zeta^{(-4)}du\int^{\infty}_0 ds\cdot s e^{-s(2{\cal W}\bar{\cal
W} +4q^-q^+)} D^+{\cal W}D^+{\cal W}\bar{D}^+\bar{\cal
W}\bar{D}^+\bar{\cal W} $$ $$\times \{1+\frac{s^4}{2\cdot 5!}
D^4{\cal W}^2 \bar{D}^4\bar{\cal W}^2 + ... \}. $$ Leading
low-energy correction corresponds to $F^4$-term. Performing the
integral over $s$ we have
\begin{equation}\label{n1}
\Gamma_{F^4}=\frac{1}{(4\pi)^2}\int d\zeta^{(-4)}du\frac{1}{16}
\frac{D^+{\cal W}D^+{\cal W}}{{\cal W}^2}\frac{\bar{D}^+\bar{\cal
W}\bar{D}^+\bar{\cal W}}{\bar{\cal
W}^2}\frac{1}{(1-(-\frac{2q^-q^+}{{\cal W}\bar{\cal W}}))^2}
\end{equation}
$$ =\sum^{\infty}_{k=0} \frac{1}{(4\pi)^2}\int d\zeta^{(-4)}du
\frac{1}{16}\frac{D^+{\cal W} ...\bar{D}^+\bar{\cal W}}{({\cal
W}\bar{\cal W})^{k+2}} (k+1)(-2q^-q^+)^k $$ $$ =
\frac{1}{(4\pi)^2}\int d\zeta^{(-4)}du
\frac{1}{16}\{D^{+2}\ln{\cal W}\bar{D}^{+2}\ln\bar{\cal W}
+\sum^{\infty}_{k=1}\frac{1}{k^2(k+1)} D^{+2}\frac{1}{{\cal
W}^k}\bar{D}^{+2}\frac{1}{\bar{\cal W}^k}(-2q^-q^+)^k\}, $$
\begin{equation}\label{n2}
\Gamma_{F^4}=\frac{1}{(4\pi)^2}\int d^{12}z  \{\ln{\cal
W}\ln\bar{\cal W} +
\sum^{\infty}_{k=1}\frac{1}{k^2(k+1)}(\frac{-2q^{ai}q_{ai}}{{\cal
W}\bar{\cal W}})^k\}~. \end{equation} That exactly coincides with
the earlier results\footnote{We use $\int
d\zeta^{(-4)}(D^+)^4=\int d^{12}z$ and $\int du=1$ since in the
central basis of the ${\cal N}=2$ harmonic superspace the
hypermultiplet superfields are expressed on shell as $q^{\pm
a}=q^{ia}u^{\pm}_i$ and ${\cal W}$ is harmonic independent.}
\cite{1}, \cite{1b}, \cite{bbp}: $$
\Gamma_{F^4}=\frac{1}{(4\pi)^2}\int d^{12}z  \{\ln{\cal
W}\ln\bar{\cal W} +  \mbox{Li}_2(X) +\ln(1-X)
-\frac{1}{X}\ln(1-X)\}, $$ where $X=\frac{-2q^{ai}q_{ai}}{{\cal
W}\bar{\cal W}}$ and $\mbox{Li}_2(X)$ is the Euler dilogarithm
function.

Next-to-leading correction corresponds to $F^8$-term\footnote{No
$F^6$ quantum correction occurs in the one-loop effective action
for ${\cal N}=4$ SYM \cite{fra}, \cite{bkt}}. We have
\begin{equation}\label{n3}
\Gamma_{F^8}=\sum_{k=0}^{\infty} \frac{1}{2(4\pi)^2}\int
d\zeta^{(-4)}du \frac{1}{16}\frac{D^+{\cal W}...\bar{D}^+\bar{\cal
W}}{({\cal W}\bar{\cal W})^6}D^4 {\cal W}^2\bar{D}^4 \bar{\cal
W}^4 \end{equation} $$ \times
\frac{1}{5!}(k+1)(k+2)(k+3)(k+4)(k+5)(\frac{-2q^-q^+}{{\cal
W}\bar{\cal W}})^k $$ Using the transformations  $$D^4 {\cal W}^2
= 4D^+_\alpha D^{-\beta} {\cal W} D^{+\alpha} D^-_\beta{\cal W},$$
and $$D^{+\alpha} D^+_\alpha (D^{+\delta} {\cal W} D^+_\delta
{\cal W} D^{-\beta} {\cal W} D^-_\beta {\cal W}) = 2
D^{+\delta}{\cal W} D^+_\delta{\cal W} D^+_\alpha D^{-\beta}{\cal
W} D^{+\alpha} D^-_\beta{\cal W},$$
we get the chain of
identities $$ \frac{1}{({\cal W})^{k+6}}D^+ {\cal W}D^+ {\cal
W}D^4 {\cal W}= 2 D^+ D^+ \{\frac{D^+ {\cal W}D^+ {\cal W}}{{\cal
W}^{k+4}}\frac{D^- {\cal W}D^- {\cal W}}{{\cal W}^2}\} $$ $$ =-2
D^+ D^+\{\frac{1}{(k+2)(k+3)} D^+ D^+\frac{1}{{\cal W}^{k+2}}
D^-D^- \ln{\cal W}\} $$ $$ =-2 D^+ D^+\{\frac{1}{(k+2)(k+3)}
\frac{1}{{\cal W}^{k+2}} D^4 \ln{\cal W}\}. $$ Similar
transformations are done for the complex conjugate term. After
restoration of the full measure $d^{12}z =d\zeta^{(-4)} (D^+)^4$
we have the factor $$\frac{1}{(k+2)^2(k+3)^2} \frac{1}{\bar{\cal
W}^2}D^4 \ln{\cal W} \frac{1}{{\cal W}^2}\bar{D}^4 \ln\bar{\cal W}
(\frac{-2q^-_a q^{+a}}{{\cal W}\bar{\cal W}})^k. $$ This allows to
write $\Gamma_{F^8}$ in the form \cite{bbp}:
\begin{equation}\label{n4}
\Gamma_{F^8}=\frac{1}{2(4\pi)^2 5!}\int d^{12}z
\sum_{k=0}^{\infty} \frac{(k+1)(k+4)(k+5)}{(k+2)(k+3)}\Psi^2
\bar{\Psi}^2 (\frac{-2q^-_a q^{+a}}{{\cal W}\bar{\cal W}})^k
\end{equation}
$$
=\frac{1}{2(4\pi)^2 5!}\int d^{12}z \Psi^2 \bar{\Psi}^2\{{1\over
(1-X)^{2}}+{4\over (1-X)}+ {6X-4\over X^{3}}\ln (1-X)+4{X-1\over
X^{2}}\}.
$$
It coincides with the result of ${\cal N}=1$ calculations derived on the base of above prescription
\cite{bbp}.

The analogous consideration allows us in principle to get any term
$\Gamma_{F^{2n}}$ of derivative expansion of the effective action
(\ref{N4}). We pay attention that the integrals over analytic
subspace are transformed to the integrals over full ${\cal N}=2$
superspace in each term of expansion.


\section{Summary}


We have studied the problem of one-loop low-energy effective
action in ${\cal N}=4$, $SU(2)$ SYM theory. The theory is
formulated in ${\cal N}=2$ harmonic superspace and possesses the
manifest off-shell ${\cal N}=2$ supersymmetry and extra on-shell
hidden ${\cal N}=2$ supersymmetry. We developed a new approach to
derivations the effective action depending on all fields of ${\cal
N}=4$ vector multiplet keeping the manifest ${\cal N}=2$
supersymmetry at all steps of calculations.

From  ${\cal N}=2$ supersymmetric point of view, the effective action
under consideration depends on fields of ${\cal N}=2$ vector
multiplet and hypermultiplet. The theory is quantized in
framework of the ${\cal N}=2$ background field method which allows us to
obtain the effective action in manifestly gauge invariant form. We
carried out the calculations of the effective action in low-energy
approximation assuming the ${\cal N}=2$ superstrengths ${\cal W},
{\bar{\cal W}}$ are on-shell and space-time independent and space-time
independent hypermultiplet superfields $q^{a}$. The effective
action is given by integral over analytic subspace of harmonic
superspace of function depending on ${\cal W}, {\bar{\cal W}}$,
their spinor covariant derivatives and hypermultiplet superfields.
This dependence is exactly found under the low-energy assumption. The
effective action obtained is an extension of the results obtained in
\cite{1}, \cite{1b} where only lowest powers of
the spinor covariant derivatives of superfield strengths have been
taken into account. We also check and justify the results of Ref.
\cite{bbp} where the one-loop effective action in hypermultiplet
sector has been found in terms of ${\cal N}=1$ superfields using
a special gauge fixing and some heuristic prescription about
reconstruction of manifestly ${\cal N}=2$ supersymmetric form of
effective action. In the given work this problem is solved
automatically.

We have developed a general method to derive the one-loop,
low-energy effective action of the theory under consideration
completely in terms of ${\cal N}=2$ harmonic superspace. Starting
point of our consideration is a series of covariant harmonic
supergraphs with arbitrary number of external hypermultiplet legs.
Each of such supergraphs is written as an integral over analytic
subspace and all contributions are summed up. The result is given
by expression (\ref{finally}). This expression is analyzed on the
base of proper-time method and techniques of operator symbols (the
basic aspects of this techniques in ${\cal N}=2$ harmonic
superspace are developed in Section 5) using the key relation
(\ref{h3}) (this new relation is derived in Appendix B). As a
result we obtain the final expression (\ref{N4}) for the effective
action.

At present there are at least two vast open problems concerning
the hypermultiplet dependence of effective action in ${\cal N}=4$
SYM theory. The first one is a problem of effective action where the
spinor covariant derivatives of hypermultiplet superfields are not
vanishing. In this case, the low-energy effective action could be
constructed in form of expansion in spinor derivative of
superstrengths and hypermultiplet superfields. In particular, such
an expansion would help to clarify an answer to the question whether or not the
effective action is invariant under the hidden on-shell ${\cal N}=2$
supersymmetry like the classical action. The second one is a
problem of hypermultiplet dependence of effective action at higher
loops. We hope that the methods developed in this work will be
useful for study both of the above problems.

\section{Acknowledgements}
The authors are grateful to A.T. Banin and S.M. Kuzenko for useful
discussions. I.L.B is grateful to Trinity College, Cambridge for
financial support. The work was supported in part by RFBR grant,
project No 03-02-16193. The work of I.L.B was also partially
supported by INTAS grant, INTAS-03-51-6346, joint RFBR-DFG grant,
project No 02-02-04002, DFG grant, project No 436 RUS 113/669,
grant for LRSS, project No 1252.2003.2. The work of N.G.P was
supported in part by RFBR grant, project No 05-02-16211.


\appendix

\renewcommand{\thesection}{Appendix \Alph{section}.}
\renewcommand{\theequation}{\Alph{section}.\arabic{equation}}

\setcounter{equation}{0}

\setcounter{section}{1}
\section*{Appendix A. Useful rep\-re\-sen\-tat\-ion of the Ba\-ker - Camp\-bell - Haus\-dorff formula}
\addcontentsline{toc}{section}{ Appendix A. Useful rep\-re\-sen\-tat\-ion of the Ba\-ker
- Camp\-bell - Haus\-dorff formula}


The heat kernel associated with the operator ${\cal O}$ is given as
matrix element of the operator $e^{{\cal O}}$. In many cases the
operator ${\cal O}$ is a linear combination of basis operators which
form some Lie (super)algebra. Then evaluation of the heat kernel
is simplified drastically by using the  Baker-Campbell-Hausdorf
(BCH) fromula\footnote{N.P. would like to thank L.L. Salcedo for
enlightening and useful discussions on various aspects of the BCH
formula}. This formula states that the exponent of the sum of two
noncommuting operators $A$ and $B$
can be presented as a series in powers of the commutator $[A,B]$.

We derive another representation of BCH-formula
\begin{equation}\label{bch}
e^{A+B}=e^{C_1+C_2+C_3+ ...}e^A,
\end{equation}
where the operators $C_k$ are expressed through the commutators of the
operators $A$ and $B$. We show here that the operators
$C_k$ can be defined in such a way that each $C_k$ has k-th power in
operator $B$ and infinite power in operator $A$. This representation
of BCH-formula (\ref{bch}) can be useful in the case when the operators
$A$ and $B$ are associated with some Lie (super)algebra which allows us to summarize
the operator series $C_1+C_2+C_3+...$ in the explicit form.

We introduce in (\ref{bch}) the variables $t$ as follows:
\begin{equation}\label{B91}
e^{A+tB}=e^{tC_1 +t^2C_2 +t^3C_3+...} e^{A}~.
\end{equation}
Next, let us define the function
\begin{equation}\label{F}
{\cal F} = e^{A+ t B}e^{-A}=
e^{\sum_{k=1}^{\infty}t^kC_k(A,B)}
\end{equation}
 and find the proper operators $C_k$ in (\ref{B91}).
It is obvious that at $t=0$ the identity (\ref{B91})
takes place and at $t=1$ ones get the initial relation (\ref{bch}).
We calculate the logarithmic derivative of the function (\ref{B91}) with respect
to $t$. On one
hand, we have \begin{equation} \dot{\cal F}{\cal F}^{-1} =\int^1_0
d\tau e^{\tau A+\tau t B} B e^{-\tau A -\tau tB}~.
\end{equation}
On the other hand \begin{equation} {\cal F}=e^{t C_1 +t^2
C_2 +...} ,\quad \dot{\cal F}{\cal F}^{-1} = C_1  + 2t C_2 +...
\end{equation}
If we put $t=0$, we find \begin{equation} C_1 =\int^1_0 d\tau
e^{\tau A}Be^{-\tau A} =\int^1_0 d\tau {\cal B}(\tau)~.
\end{equation} In order to obtain $C_2$ we have to calculate first
derivative with respect $t$ of the logarithmic derivation:
\begin{equation}\label{B4}
\frac{d}{d t}(\dot{\cal F}{\cal
F}^{-1})=\int^1_0 \int^1_0 d \tau' d \tau \{e^{\tau' \tau A +\tau'
\tau t B}(\tau B)e^{ -\tau' \tau A -\tau' \tau t B}\}\{e^{\tau A +
\tau t B} B e^{ -\tau A - \tau t B}\}
\end{equation}
$$ + \int^1_0 \int^1_0 d\tau' d\tau \{e^{\tau A +\tau t B}B
e^{-\tau A -\tau t B}\} \{e^{\tau' \tau A +\tau' \tau t B}(-\tau
B)e^{ -\tau' \tau A -\tau' \tau t B}\}~. $$
At $t=0$ we get
the expression for $C_2$:
\begin{equation}
2C_2 =\int^1_0 \int^1_0
d\tau d\tau' \cdot \tau \cdot[{\cal B}(\tau'\tau), {\cal
B}(\tau)]~.
\end{equation}
Following  the same way one can obtain all operators
$C_k$. For example, to get the operator
$C_3$ we have to calculate second  derivation of above logarithmic
derivative. It leads to
\begin{equation}
6C_3 +[C_1,C_2]=\int^1_0
d\tau'' d\tau' d\tau \cdot \tau^2 \{[{\cal B}(\tau' \tau),[{\cal
B}(\tau'' \tau), {\cal B}(\tau)]] - \tau'[{\cal B}(\tau),[{\cal
B}(\tau'' \tau' \tau),{\cal B}(\tau' \tau)]]\}~,
\end{equation}
where $C_1$ and $C_2$ have been obtained above. The operator $C_4$ is found from
\begin{equation}
24C_4 +6 [C_1,C_3]+ [C_1,[C_1,C_2]] = \int^1_0 d\tau''' d\tau''
d\tau' d\tau
\end{equation}
$$\times \lceil \tau^2 \{\tau' \tau [[{\cal B}(\tau''' \tau'
\tau), {\cal B}(\tau' \tau ],[{\cal B}(\tau'' \tau),{\cal
B}(\tau)]] -\tau'' \tau [{\cal B}(\tau' \tau)[{\cal B}(\tau)
[{\cal B}(\tau''' \tau'' \tau),{\cal B}(\tau'' \tau)] $$ $$+\tau
[{\cal B}(\tau' \tau)[{\cal B}(\tau'' \tau)[{\cal B}(\tau'''
\tau), {\cal B}(\tau)]]]\} $$ $$ \tau^2 \tau'\{\tau'' \tau' \tau
[[[{\cal B}(\tau''' \tau'' \tau' ),{\cal B}(\tau'' \tau'
\tau)]{\cal B}(\tau' \tau)]{\cal B}(\tau)] +\tau' \tau [[[{\cal
B}(\tau'' \tau' \tau)[{\cal B}(\tau''' \tau' \tau),{\cal B}(\tau'
\tau)]{\cal B}(\tau)] + $$ $$\tau [[{\cal B}(\tau'' \tau'
\tau),{\cal B}(\tau' \tau )],[{\cal B}(\tau''' \tau),{\cal
B}(\tau)]]\}\rfloor $$ and etc. Following the same procedure we
can obtain, in principle, all terms of the BCH series in
(\ref{bch}). Note that when Lie (super)algebra under consideration
has a central charge, this series terminates at some finite
order and the representation (\ref{bch}) takes a simple form.

In the case under consideration, the problem is in rewriting
the exponent (\ref{h3}) of sum of the operators satisfying the
algebraic relations (\ref{salg1}), (\ref{salg3}) as a product of
the exponents of individual operators. This problem is solved in
two steps. First, we take A in (\ref{bch}) in the form $A^+D^-
+\bar{A}^+ \bar{D}^-$ and obtain $C_1$ as a linear combination
of operators $\frac{1}{2}{\cal D}^{\alpha\dot\alpha}{\cal
D}_{\alpha\dot\alpha}$, $f^{\alpha\dot\alpha}( A^{\pm},
\bar{A}^{\pm}, N, \bar{N}) {\cal D}_{\alpha\dot\alpha}$ with some
definite coefficients\footnote{indeed, e.g. $[A^+{\cal
D}^-,\frac{1}{2}{\cal D}^{\alpha\dot\alpha}{\cal
D}_{\alpha\dot\alpha}]=-A^{+\alpha}\bar{A}^{-\dot\alpha}{\cal
D}_{\alpha\dot\alpha}$} plus some definite function of arguments
${\cal W}, \bar{\cal W}, A^{\pm}, \bar{A}^{\pm}, N, \bar{N}$ as
the central element. This means, that all other operators $C_2, C_3,
...$ will be proportional to the operator ${\cal
D}_{\alpha\dot\alpha}$ with some coefficient functions. Therefore,
the series $C_1 + C_2+ C_3 + ....$ is reduced to summing up these
coefficients functions what can be done in explicit form. The
result is $-s\frac{1}{2}{\cal D}^{\alpha\dot\alpha}{\cal
D}_{\alpha\dot\alpha}$ plus $f^{\alpha\dot\alpha}( A^{\pm},
\bar{A}^{\pm}, N, \bar{N}) {\cal D}_{\alpha\dot\alpha}$ plus some
central element. Second, we apply again the formula (\ref{bch}) to
the expression $exp(-s\frac{1}{2}{\cal D}^{\alpha\dot\alpha}{\cal
D}_{\alpha\dot\alpha} + f^{\alpha\dot\alpha}{\cal
D}_{\alpha\dot\alpha})$ and take $A$ as $-s\frac{1}{2}{\cal
D}^{\alpha\dot\alpha}{\cal D}_{\alpha\dot\alpha}$. All operators
$C_k$ will be again proportional to the single operator ${\cal
D}_{\alpha\dot{\alpha}}$ and series of coefficient functions can
be summed up in explicit form. As a result we get the right had
side of expression (\ref{h3}). All coefficient functions are given
in Appendix B. These coefficient functions can also be found from
differential equations (\ref{h4}), (\ref{h5}). The results
obtained from both BCH-formula and the above differential
equations coincide.

\setcounter{equation}{0}

\setcounter{section}{2}

\section*{Appendix B. Coefficient functions in decomposition of heat kernel}
\addcontentsline{toc}{section}{Appendix B. Coefficient functions
in decomposition of heat kernel}

The linear differential equations (\ref{h4}) and (\ref{h5}) for
$f_{\alpha\dot\alpha}(s)$ and $\Omega(s)$ can be solved exactly in
the form (\ref{h6}), (\ref{h8}). Coefficients of the power
expansion $f_{\alpha\dot\alpha}(s)$ over basis of Grassmann
elements $A^+_{\alpha}, \bar{A}^+_{\dot\alpha}$ are given as
follows:
\begin{equation}\label{h7}
{\cal N}^{\delta\dot\delta}_{\alpha\dot\alpha} = \int^s_0 d\tau
(\frac{e^{-\tau N}-e^{\tau\bar{N}}}{N+\bar{N}} \cdot
e^{-s\bar{N}})^{\delta\dot\delta}_{\alpha\dot\alpha}
=-\frac{e^{-sF}-1}{N F}+ \frac{e^{-s\bar{N}}-1}{N\bar{N}},
\end{equation}

\begin{equation}\label{h7a}
\bar{\cal N}^{\dot\delta\delta}_{\dot\alpha\alpha} =\int^s_0 d\tau
(\frac{e^{-\tau \bar{N}}-e^{\tau N}}{N +\bar{N}} \cdot
e^{-sN})^{\dot\delta\delta}_{\dot\alpha\alpha}
=-\frac{e^{-sF}-1}{\bar{N}F} +\frac{e^{-sN}-1}{N\bar{N}}.
\end{equation}

Coefficients of the power expansion $\Omega(s)$ over basis of
Grassmann elements $A^+_{\alpha}, \bar{A}^+_{\dot\alpha}$ are given
as follow:

\begin{equation}\label{h9}
\Omega^-_\alpha  = -\bar{\cal
W}\{\frac{e^{-sN}+sN-1}{N^2}\}^\beta_\alpha A^-_\beta,
\end{equation}
\begin{equation}\label{h10}
\bar{\Omega}^-_{\dot\alpha}=-{\cal W}
\bar{A}^-_{\dot\beta}\{\frac{e^{-s\bar{N}} +
s\bar{N}-1}{\bar{N}^2}\}^{\dot\beta}_{\dot\alpha},
\end{equation}
\begin{equation}\label{h11}
\Psi^{(-2)}=\frac{1}{8}(\bar{A}^-)^2
\mbox{tr}\sum^{\infty}_{n=0}\sum^n_{p=1}\frac{s^{n+2}}{(n+2)!}C_n^p(-F)^{n-p}\{N^{p-1}-(-1)^n
\bar{N}^{p-1} \}
\end{equation}
$$ =(\bar{A}^-)^2 \{\frac{s^3}{6} + \frac{s^5}{5!}(N^2+\bar{N}^2)
+...\}, $$
\begin{equation}\label{h12}
\bar{\Psi}^{(-2)}=\frac{1}{8}({A}^-)^2
\mbox{tr}\sum^{\infty}_{n=0}\sum^n_{p=1}\frac{s^{n+2}}{(n+2)!}C_n^p(-F)^{n-p}\{\bar{N}^{p-1}-(-1)^n
N^{p-1} \}
\end{equation}
$$ =(A^-)^2 \{\frac{s^3}{6} +\frac{s^5}{5!}(N^2 +\bar{N}^2)+...\},
$$ $$ \Psi^{\dot\alpha(-2)}_\alpha = \Psi^{\dot\alpha
\delta}_{\alpha\dot\delta} A^-_\delta \bar{A}^{-\dot\delta}, $$
\begin{equation}\label{h13}
\Psi^{\dot\alpha \delta}_{\alpha\dot\delta}=
\frac{1}{N\bar{N}(N-\bar{N})}
+\frac{1}{N\bar{N}}\{\frac{e^{-sN}}{\bar{N}}-
\frac{e^{s\bar{N}}}{N}\}
+\frac{N^2+\bar{N}^2}{2N^2\bar{N}^2(\bar{N}-N)}e^{s(\bar{N}-N)}
+\frac{\bar{N}^2e^{sF}-N^2 e^{-sF}}{2N^2\bar{N}^2(N+\bar{N})}
\end{equation}
$$ =\frac{s^3}{3} + \frac{s^4}{8}(\bar{N}-N)
+\frac{7s^5}{5!}(N^2+\bar{N}^2)+...~. $$

Coefficients of the derivative expansion $\exp\{-s(A^+{\cal D}^-
+\bar{A}^+\bar{\cal D}^-)\}$ defined in (\ref{h14}) are given as
follows:

\begin{equation}\label{h15}
a^{+\alpha}=A^{+\beta}(\frac{e^{-sN}-1}{N})^\alpha_\beta, \quad
\bar{a}^{+\dot\alpha}=\bar{A}^{+\dot\beta}(\frac{e^{-s\bar{N}}-1}{\bar{N}})^{\dot\alpha}_{\dot\beta}~,
\end{equation}

\begin{equation}\label{h17}
f^{+2}=-\frac{1}{4}(A^+)^2\mbox{tr}(\frac{\cosh(sN)-1}{N^2}),
\quad
\bar{f}^{+2}=-\frac{1}{4}(\bar{A}^+)^2\mbox{tr}(\frac{\cosh(s\bar{N})-1}{\bar{N}^2}),
\end{equation}
\begin{equation}\label{h19}
f^{+2\dot\alpha\alpha}=-A^{+\beta}\bar{A}^{+\dot\beta}(\frac{e^{-sN}-1}{N})^\alpha_\beta
(\frac{e^{-s\bar{N}}-1}{\bar{N}})^{\dot\alpha}_{\dot\beta}~,
\end{equation}

\begin{equation}\label{h20}
\bar{\Xi}^{+3\dot\alpha}=-\frac{1}{4}(A^+)^2\bar{A}^{+\dot\beta}(\frac{e^{-s\bar{N}}-1}{\bar{N}})^{\dot\alpha}_{\dot\beta}
\mbox{tr}(\frac{\cosh(sN)-1}{N^2})~,
\end{equation}

\begin{equation}\label{h21}
\Xi^{+3\alpha}=-\frac{1}{4}(\bar{A}^+)^2A^{+\beta}(\frac{e^{-sN}-1}{N})^\alpha_\beta
\mbox{tr}(\frac{\cosh(s\bar{N}-1)}{\bar{N}^2})~,
\end{equation}
\begin{equation}\label{h22}
\Omega^{+4}= -\frac{1}{16} (A^+)^2(\bar{A}^+)^2
\mbox{tr}(\frac{\cosh(s{N})-1}{{N}^2})
\mbox{tr}(\frac{\cosh(s\bar{N})-1}{\bar{N}^2})~.
\end{equation}


\begin{thebibliography}{000}

\addcontentsline{toc}{section}{References}

\bibitem{T}W. Taylor, Lectures on D-branes, Gauge Theory and M(atrices), [hep-th/9801182].

\bibitem{HF}E. D'Hoker, D.Z. Freedman, Supersymmetric Gauge Theories and the AdS/CFT
Correspondence, TASSI 2001 Lecture Notes, [hep-th/0201253].

\bibitem{M}J.M. Maldacena, The large N limit of superconformal
field theories and supergravity, {\it Adv. Theor. Math. Phys.}
{\bf 2} 231 (1998) [hep-th/9711200]; S.S. Gubser, I.R. Klebanov
and A.M. Polyakov, Gauge theory correlators from non-critical
string theory, {\it Phys. Lett.} {\bf B428} 105 (1998) [hep-th/
9802109]; E. Witten, Anti-de Sitter space and holography, {\it
Adv. Theor. Math. Phys.} {\bf 2} 253 (1998) [hep-th/9802150]; O.
Aharony, S.S. Gubser, J. Maldacena, H. Ooguri, Y. Oz, Large N
field theories, string theory and gravity, {\it Phys. Repts.} {\bf
332} (2000) 163 [hep-th/9905111].

\bibitem{chep}I. Chepelev, A.A. Tseytlin, Long-distance interactions of branes:
correspondence between supergravity and super Yang-Mills
descriptions, {\it Nucl.Phys.} {\bf B515} (1998) 73
[hep-th/9709087]; A.A. Tseytlin, ``Born-Infeld action,
super\-symmetry and string the\-ory'', in: Yuri Golfand memorial
volume, ed. M. Shifman, World Scientific, 2000, pp. 417-452~
[hep-th/9908105].

\bibitem{bkt}I.L. Buchbinder,
S.M. Kuzenko, A.A. Tseytlin, On Low-Energy Effective Actions in N
= 2, 4 Superconformal Theories in Four Dimensions, {\it Phys.
Rev.} {\bf D62} (2000) 045001 [hep-th/9911221].

\bibitem{bpt}I.L. Buchbinder, A.Yu. Petrov,
A.A. Tseytlin, Two-loop N=4 Super Yang Mills effective action and
interaction between D3-branes, {\it Nucl. Phys.} {\bf B621} (2002)
179 [hep-th/0110173].

\bibitem{gikos1}A. Galperin, E. Ivanov, S. Kalitzin, V. Ogievetsky
and E. Sokatchev, N=3 supersymmetric gauge theory, {\it Phys.
Lett.} {\bf B151} (1985) 215; Unconstrained off-shell N=3
supersymmetric Yang-Mills theory, {\it Class. Quant. Grav.} {\bf
2} (1985) 155; A. Galperin, E. Ivanov, V. Ogievetsky, Superspaces
for $N=3$ supersymmetry, {\it Sov. J. Nucl. Phys.} {\bf 46} (1987)
543.

\bibitem{gios}A.S. Galperin, E.A. Ivanov, V.I. Ogievetsky and E.S Sokatchev,
{\it Harmonic Superspace}, Cambridge Univ. Press, (2001) 306 p.

\bibitem{del}F. Delduc, J. McCabe, The quantization of N=3 super-Yang-Mills
off-shell in superspace, {\it Class. Quantum Grav.} {\bf 6} (1989)
233.

\bibitem{bisz} I.L. Buchbinder, E.A. Ivanov, I.B. Samsonov, B.M.
Zupnik, Scale Invariant Low-Energy Effective Action in ${\cal
N}=3$ SYM Theory, {\it Nucl.Phys.} {\bf B689} (2004) 91
[hep-th/0403053].

\bibitem{iz}E.A. Ivanov, B.M. Zupnik, N=3 supersymmetric
Born-Infeld theory, {\it Nucl.Phys.} {\bf B618} (2001) 3
[hep-th/0110074]; E. Ivanov, Towards higher-N superextensions of
Born-Infeld theory, {\it Russ. Phys. J.} {\bf 45}, 697 (2002)
[hep-th/0202201].

\bibitem{gikos}A. Galperin, E. Ivanov, S. Kalitzin, V. Ogievetsky
and E. Sokatchev, Unconstrained N=2 matter, Yang-Mills and
supergravity theories in harmonic superspace, {\it Class. Quant.
Grav.} {\bf 1} (1984) 469; A. Galperin, E. Ivanov, V. Ogievetsky
and E. Sokatchev, Harmonic supergraphs. Green functions, {\it
Class. Quant. Grav.} {\bf 2} (1985) 601; Harmonic supergraphs.
Feynman rules and examples, {\it Class. Quant. Grav.} {\bf 2}
(1985) 617.

\bibitem{oh}N. Ohta, H. Yamaguchi, Superfield perturbation theory in harmonic
superspace, {\it Phys. Rev.} {\bf D32} (1985) 1954.

\bibitem{backgr}I.L.
Buchbinder, E.I. Buchbinder, S.M. Kuzenko, B.A. Ovrut, The
Background Field Method for N = 2 Super Yang-Mills Theories in
Harmonic Superspace, {\it Phys. Lett.} {\bf B417} (1998)  61
[hep-th/9704214]; I. Buchbinder, S. Kuzenko, B. Ovrut, Covariant
Harmonic Supergraphity for N = 2 Super Yang--Mills Theories, in J.
Wess and E. Ivanov (Eds.), {\it Supersymmetries and Quantum
Symmetries}, Springer, Berlin (1999) [hep-th/9810040].

\bibitem{bbiko} E.I. Buchbinder, I.L. Buchbinder, E.A. Ivanov, S.M. Kuzenko,
B.A. Ovrut, Low-Energy Effective Action in N = 2 Supersymmetric
Field Theories, {\it Physics of Particles and Nuclei}, {\bf 32}
641 (2001).

\bibitem{2loop}S.M. Kuzenko, I.N. McArthur, On the Background Field Method Beyond One Loop:
A manifestly covariant derivative expansion in super Yang-Mills
theories, {\it JHEP} {\bf 0305} (2003) 015 [hep-th/0302205];
Low-energy dynamics in N = 2 super QED: Two-loop approximation,
{\it JHEP} {\bf 0310} (2003) 029 [hep-th/0308136]; On the two-loop
four-derivative quantum corrections in 4D N = 2 superconformal
field theories, {\it Nucl.Phys.} {\bf B683} (2004) 3
[hep-th/0310025]; Relaxed super self-duality and effective action,
{\it Phys. Lett.} {\bf B591} (2004) 304, [hep-th/0403082]; Relaxed
super self-duality and N = 4 SYM at two loops, {\it Nucl.Phys.}
{\bf B697} (2004) 89 [hep-th/0403240].


\bibitem{howe}P.J. Heslop, P.S. Howe, Aspects of $N=4$ SYM,
{\it JHEP} {\bf 0401} (2004) 058 [hep-th/0307210]; J. M. Drummond,
P. J. Heslop, P. S. Howe, S. F. Kerstan, Integral invariants in
N=4 SYM and the effective action for coincident D-branes, {\it
JHEP} {\bf 0308} (2003) 016 [hep-th/0305202]; P.C. Argyres, A. M.
Awad, G. A. Braun, F.P. Esposito, Higher-Derivative Terms in N=2
Supersymmetric Effective Actions, {\it JHEP} {\bf 0307} (2003) 06~
[hep-th/0306118].

\bibitem{ARG}P.C. Argyres, M.R. Plesser, N. Seiberg, The Moduli Space of $N=2$ SUSY QCD and
Duality in $N=1$ SUSY QCD, {\it Nucl.Phys.} {\bf B471} (1996) 159
[hep-th/9603042].


\bibitem{6a} M. Dine, N. Seiberg, Comments on Higher Derivative Operators in Some SUSY Field
Theories, {\it Phys.Lett.} {\bf B409} 239 (1997) [hep-th/9705057];
D.A. Lowe, R. von Unge, Constraints on Higher Derivative Operators
in Maximally Supersymmetric Gauge Theory, {\it JHEP} {\bf 9811}
014 (1998) [hep-th/9811017]; F. Gonzalez-Rey, M. Ro\v{c}ek,
Nonholomorphic N=2 terms in N=4 SYM: 1-Loop Calculation in N=2
superspace, {\it Phys. Lett.} {\bf B434} 303 (1998)
[hep-th/9804010]; V. Periwal, R. von Unge, Accelerating D-branes,
{\it Phys. Lett.} {\bf B430} 71 (1998) [hep-th/9801121].

\bibitem{bbku}I.L. Buchbinder, S.M. Kuzenko, Comments on the Background Field Method in Harmonic Superspace: Non-holomorphic Corrections in N=4
SYM, {\it Mod. Phys. Lett.} {\bf A13} (1998)  1623
[hep-th/9804168]; E.L. Buchbinder, I.L. Buchbinder and S.M.
Kuzenko, Non-holomorphic effective potential in N = 4 SU(n) SYM,
{\it Phys.Lett.} {\bf B446} (1999) 216 [hep-th/9810239].

\bibitem {nonr}I.L. Buchbinder, S.M. Kuzenko, B.A. Ovrut, On the D = 4, N = 2 Non-Renormalization
Theorem, {\it Phys. Lett.} {\bf B433} 335 (1998) [hep-th/9710142];
I.L. Buchbinder, A.Yu. Petrov, N=4 Super Yang-Mills Low-Energy
Effective Action at Three and Four Loops, {\it Phys.Lett.} {\bf
B482} (2000) 429 [hep-th/0003265].

\bibitem{dwgr} B. de Wit, M.T. Grisaru, M. Ro\v{c}ek, Nonholomorphic Corrections to the One-Loop N=2 Super Yang-Mills
Action, {\it Phys. Lett.} {\bf B374} 297 (1996) [hep-th/9601115];
A.De Giovanni, M.T. Grisaru, M. Ro\v{c}ek, R. von Unge, D.Zanon,
The N=2 Super Yang-Mills Low-Energy Effective Action at Two Loops,
{\it Phys. Lett.} {\bf B409} 251 (1997) [hep-th/9706013]; A. Yung,
Higher Derivative Terms in the Effective Action of N=2 SUSY QCD
from Instantons, {\it Nucl. Phys.} {\bf B512} (1998) 79
[hep-th/9705181]; D. Bellisai, F. Fucito, M. Matone, G.
Travaglini, Non-holomorphic terms in N=2 SUSY Wilsonian actions
and RG equation, {\it Phys. Rev.} {\bf
 D56} (1997) 5218 [hep-th/9706099]; N.  Dorey, V.V. Khoze, M.P.
Mattis, M.J. Slater, W.A. Weir, Instantons, Higher-Derivative
Terms, and Nonrenormalization Theorems in Supersymmetric Gauge
Theories, {\it Phys. Lett.} {\bf B408} (1997) 213
[hep-th/9706007].

\bibitem{Bec}K. Becker, M. Becker, A Two-Loop Test of M(atrix) Theory,
{\it Nucl. Phys.} {\bf B506} (1997) 48 [hep-th/9705091]; K.
Becker, Testing M(atrix)-Theory at Two Loops, {\it Nucl. Phys.
Proc. Suppl.} {\bf 68} (1998) 165 [hep-th/9709038]; S. Sethi,
Structure in Supersymmetric Yang-Mills Theory, {\it JHEP} {\bf
0410} (2004) 001 [hep-th/0404056].


\bibitem{1}I.L. Buchbinder, E.A.
Ivanov, Complete N=4 Structure of Low-Energy Effective Action in
N=4 Super Yang-Mills Theories, {\it Phys. Lett.} {\bf B524} (2002)
208 [hep-th/0111062].

\bibitem{1b}I.L. Buchbinder, E.A.
Ivanov, A.Yu. Petrov, Complete Low-Energy Effective action in N=4
SYM: a Direct N=2 Supergraph Calculation, {\it Nucl. Phys.} {\bf
B653} (2003) 64 [hep-th/0210241].

\bibitem{garg}I.N. McArthur, T.D. Gargett, ''Gaussian'' Approach
to Computing Supersymmetric Effective Actions, {\it Nucl.Phys.}
{\bf B494}, 525 (1997) [hep-th/9705200].

\bibitem{bbp} A.T. Banin, I.L. Buchbinder, N.G. Pletnev, One-loop effective action for ${\cal N}=4$ SYM theory in the
hypermultiplet sector - leading low-energy approximation and
beyond, {\it Phys. Rev.} {\bf D68} (2003) 065024 [hep-th/0304046];
A.T. Banin, N.G. Pletnev, On the construction of ${\cal N}=4$ SYM
effective action beyond leading low-energy approximation,
[hep-th/0401006]; Low-energy next-to-leading contributions to the
effective action in ${\cal N}=4$ SYM theory, [hep-th/0401007].


\bibitem{pb} N.G. Pletnev, A.T. Banin, Covariant technique of derivative
expansion of one-loop effective action, {\it Phys.Rev.} {\bf D60}
(1999) 105017 [hep-th/9811031]; A.T. Banin, I.L. Buchbinder, N.G.
Pletnev, Low-Energy Effective Action of N=2 Gauge Multiplet
Induced by Hypermultiplet Matter, {\it Nucl.Phys.} {\bf B598}
(2001) 371 [hep-th/0008167]; On Low-Energy Effective Action in N=2
Super Yang-Mills Theories on Non-Abelian Background, {\it
Phys.Rev.} {\bf D66} (2002) 045021 [hep-th/0205034]; Chiral
effective potential in ${\cal N}={1/2}$ non-commutative
Wess-Zumino model, {\it JHEP} {\bf 0407} (2004) 011
[hep-th/0405063].

\bibitem{singul}A. Galperin, Nguyen Anh Ky and E. Sokatchev, Coinciding
harmonic singularities in harmonic supergraphs, {\it Mod. Phys.
Lett.} {\bf A2} (1987) 33.

\bibitem{town}P.S. Howe, K.S. Stelle, P.K. Townsend,  Miraculous Ultraviolet Cancellations in Supersymmetry Made
Manifest, {\it Nucl. Phys.} {\bf B236} (1984) 125.

\bibitem{exprop} S.M. Kuzenko,  Exact propagators in harmonic
superspace, {\it Phys.Lett.} {\bf B600} (2004) 163
[hep-th/0407242].

\bibitem{24} J. Gates, M. Grisaru, M. Ro\v{c}ek and W. Siegel,
{\it Superspace or One Thousand and One Lessons in Supersymmetry},
Benjamin/Cummings, Reading MA (1983) [hep-th/0108200].

\bibitem{25} I.L. Buchbinder, S.M. Kuzenko, {\it Ideas and Methods of Supersymmetry
and Supergravity or a Walk Through Superspace}, Bristol, UK: IOP
(1998).

\bibitem{2} S.M. Kuzenko, I.N. McArthur, Effective action of ${\cal
N}=4$ super Yang-Mills: ${\cal N}=2$ superspace approach, {\it
Phys. Lett}. {\bf B506} (2001) 140 [hep-th/0101127];
Hypermultiplet effective action: ${\cal N}=2$ superspace approach,
{\it Phys.Lett}. {\bf B513} (2001) 213 [hep-th/0105121].

\bibitem{fra}E.S. Fradkin and A.A. Tseytlin,  Quantization And Dimensional Reduction: One Loop Results For
Super Yang-Mills And Supergravities In $D \geq 4$, {\it
Phys.Lett.} {\bf B123} (1983) 231;  Quantum properties of higher
dimensional and dimensionally reduced supersymmetric theories,
{\it Nucl. Phys.} {\bf B277} (1983) 252.




\end{thebibliography}
\end{document}